\documentclass[twocolumn,amsmath,amssymb,prd]{revtex4}

\def\al{\alpha}
\def\be{\beta}
\def\ga{\gamma}
\def\de{\delta}
\def\ep{\epsilon}

\def\et{\eta}
\def\th{\theta}

\def\io{\iota}
\def\vka{\varkappa}
\def\ka{\kappa}
\def\la{\lambda}

\def\rh{\rho}

\def\si{\sigma}

\def\ta{\tau}

\def\ph{\phi}

\def\ch{\chi}
\def\ps{\psi}
\def\om{\omega}
\def\Ga{\Gamma}
\def\De{\Delta}

\def\La{\Lambda}
\def\Si{\Sigma}
\def\Up{\Upsilon}

\def\cl{{\cal L}}

\def\cR{{\cal R}}

\def\mn{{\mu\nu}}

\def\fr#1#2{{{#1}\over{#2}}}
\def\frac#1#2{{\textstyle{{#1}\over{#2}}}}
\def\half{{\textstyle{1\over 2}}}
\def\ol{\overline}
\def\prt{\partial}
\def\pt{\phantom}

\def\lsim{\mathrel{\rlap{\lower4pt\hbox{\hskip1pt$\sim$}}
    \raise1pt\hbox{$<$}}}
\def\gsim{\mathrel{\rlap{\lower4pt\hbox{\hskip1pt$\sim$}}
    \raise1pt\hbox{$>$}}}

\def\etal{{\it et al.}}

\def\vev#1{\langle {#1}\rangle}

\def\tr{{\rm tr}}

\def\eff{{\rm eff}}

\def\sqr#1#2{{\vcenter{\vbox{\hrule height.#2pt
         \hbox{\vrule width.#2pt height#1pt \kern#1pt
         \vrule width.#2pt}
         \hrule height.#2pt}}}}

\newcommand{\beq}{\begin{equation}}
\newcommand{\eeq}{\end{equation}}
\newcommand{\bea}{\begin{eqnarray}}
\newcommand{\eea}{\end{eqnarray}}
\newcommand{\rf}[1]{(\ref{#1})}

\newcommand{\bM}{\begin{pmatrix}}
\newcommand{\eM}{\end{pmatrix}}

\def\nn{\nonumber}

\def\psb{\ol\ps{}}

\def\bsi{\ol\si}

\def\mbf#1{\boldsymbol #1}

\def\syjm#1#2{{}_{#1}Y_{#2}}

\def\Q{\mathcal Q}
\def\S{\mathcal S}
\def\P{\mathcal P}
\def\V{\mathcal V}
\def\A{\mathcal A}
\def\T{\mathcal T}

\def\E{\mathcal E}
\def\B{\mathcal B}
\def\K{\mathcal K}

\def\cEvec{\mbf\E}
\def\cBvec{\mbf\B}

\def\eom{E}
\def\eomz{{E_0}}
\def\BD{\Up}
\def\nv{N}

\def\sv{\Si}
\def\svvec{\mbf\Si}

\def\pvec{\mbf p}

\def\kvec{\mbf k}
\def\nvec{\mbf \nv}
\def\vvec{\mbf v}
\def\xvec{\mbf x}
\def\Pvec{\mbf P}
\def\gavec{\mbf\ga}

\def\sivec{\mbf\si}

\def\del{\mbf\nabla}

\def\Svec{\mbf S}

\def\pmag{|\pvec|}
\def\komag{|{\mbf k_0}|}
\def\krmag#1{|{\mbf k_r}(#1)|}

\def\punit{\hat p}

\def\xunit{\hat x}
\def\yunit{\hat y}
\def\zunit{\hat z}
\def\epunit{\hat\ep}
\def\thunit{\hat\th}
\def\phunit{\hat\ph}

\def\phat{\mbf\punit}

\def\xhat{\mbf\xunit}
\def\yhat{\mbf\yunit}
\def\zhat{\mbf\zunit}
\def\ephat{\mbf\epunit}
\def\thhat{\mbf\thunit}
\def\phhat{\mbf\phunit}

\def\bc#1#2{\left(\begin{smallmatrix} #1 \\ #2 \end{smallmatrix}\right)}

\def\Qhat{\widehat\Q}
\def\Shat{\widehat\S}
\def\Phat{\widehat\P}
\def\Vhat{\widehat\V}
\def\Ahat{\widehat\A}
\def\That{\widehat\T}
\def\Tdual{\widetilde{\widehat\T}\phantom{}}

\def\mhat{\widehat m}
\def\mfivehat{\widehat m_5}
\def\ahat{\widehat a}
\def\bhat{\widehat b}
\def\chat{\widehat c}
\def\dhat{\widehat d}
\def\ehat{\widehat e}
\def\fhat{\widehat f}
\def\ghat{\widehat g}
\def\Hhat{\widehat H}
\def\gdual{\widetilde{\widehat g}\phantom{}}
\def\Hdual{\widetilde{\widehat H}\phantom{}}
\def\gt{\widetilde g}
\def\Ht{\widetilde H}

\def\X{X}
\def\Y{Y}
\def\Z{Z}
\def\Xhat{\widehat\X}
\def\Yhat{\widehat\Y}
\def\Zhat{\widehat\Z}

\def\template#1#2#3#4{#1^{(#2)#4}_{#3}}
\def\acoef#1#2{\template{a}{#1}{#2}{}}

\def\ccoef#1#2{\template{c}{#1}{#2}{}}

\def\gzBcoef#1#2{\template{g}{#1}{#2}{(0B)}}
\def\goBcoef#1#2{\template{g}{#1}{#2}{(1B)}}
\def\goEcoef#1#2{\template{g}{#1}{#2}{(1E)}}
\def\HzBcoef#1#2{\template{H}{#1}{#2}{(0B)}}
\def\HoBcoef#1#2{\template{H}{#1}{#2}{(1B)}}
\def\HoEcoef#1#2{\template{H}{#1}{#2}{(1E)}}
\def\Template#1#2#3#4{(#1^{(#2)})^{(#3)}_{#4}}
\def\Edjm#1#2#3{\Template{\E}{#2}{#1}{#3}}
\def\Bdjm#1#2#3{\Template{\B}{#2}{#1}{#3}}
\def\nr{{\rm NR}}
\def\nrtemplate#1#2#3{#1^{\nr#3}_{#2}}
\def\anr#1{\nrtemplate{a}{#1}{}}

\def\cnr#1{\nrtemplate{c}{#1}{}}

\def\gzBnr#1{\nrtemplate{g}{#1}{(0B)}}
\def\goBnr#1{\nrtemplate{g}{#1}{(1B)}}
\def\goEnr#1{\nrtemplate{g}{#1}{(1E)}}
\def\HzBnr#1{\nrtemplate{H}{#1}{(0B)}}
\def\HoBnr#1{\nrtemplate{H}{#1}{(1B)}}
\def\HoEnr#1{\nrtemplate{H}{#1}{(1E)}}
\def\ur{{\rm UR}}
\def\urtemplate#1#2#3#4{#1^{\ur(#2)#4}_{#3}}
\def\aur#1#2{\urtemplate{a}{#1}{#2}{}}

\def\cur#1#2{\urtemplate{c}{#1}{#2}{}}

\def\gzBur#1#2{\urtemplate{g}{#1}{#2}{(0B)}}
\def\goBur#1#2{\urtemplate{g}{#1}{#2}{(1B)}}
\def\goEur#1#2{\urtemplate{g}{#1}{#2}{(1E)}}
\def\HzBur#1#2{\urtemplate{H}{#1}{#2}{(0B)}}
\def\HoBur#1#2{\urtemplate{H}{#1}{#2}{(1B)}}
\def\HoEur#1#2{\urtemplate{H}{#1}{#2}{(1E)}}
\def\ring#1{{\mathaccent'27 #1}}

\def\fctemplate#1#2#3{\ring{#1}^{(#2)}_{#3}}
\def\afc#1#2{\fctemplate{a}{#1}{#2}}

\def\cfc#1#2{\fctemplate{c}{#1}{#2}}

\def\gfc#1#2{\fctemplate{g}{#1}{#2}}
\def\Hfc#1#2{\fctemplate{H}{#1}{#2}}
\def\nrfctemplate#1#2{\nrtemplate{\ring{#1}}{#2}{}}
\def\anrfc#1{\nrfctemplate{a}{#1}}

\def\cnrfc#1{\nrfctemplate{c}{#1}}

\def\gnrfc#1{\nrfctemplate{g}{#1}}
\def\Hnrfc#1{\nrfctemplate{H}{#1}}
\def\urfctemplate#1#2#3{\urtemplate{\ring{#1}}{#2}{#3}{}}
\def\aurfc#1#2{\urfctemplate{a}{#1}{#2}}

\def\curfc#1#2{\urfctemplate{c}{#1}{#2}}

\def\gurfc#1#2{\urfctemplate{g}{#1}{#2}}
\def\Hurfc#1#2{\urfctemplate{H}{#1}{#2}}
\def\Kcoef#1#2{\template{\K}{#1}{#2}{}}

\def\Kur#1#2{\urtemplate{\K}{#1}{#2}{}}

\def\aclock#1#2{(\ring a_c)^{(#1)}_{#2}}
\def\cclock#1#2{(\ring c_c)^{(#1)}_{#2}}
\def\arod#1#2{(\ring a_r)^{(#1)}_{#2}}
\def\crod#1#2{(\ring c_r)^{(#1)}_{#2}}

\def\m{m_\ps}

\begin{document}
\title{Fermions with Lorentz-violating operators 
of arbitrary dimension}

\author{V.\ Alan Kosteleck\'y$^1$ and Matthew Mewes$^2$}

\affiliation{$^1$Physics Department, Indiana University, 
Bloomington, Indiana 47405, USA\\
$^2$Physics Department, Swarthmore College,
Swarthmore, Pennsylvania 19081, USA}

\date{IUHET 577, August 2013}

\begin{abstract}

The theoretical description of fermions 
in the presence of Lorentz and CPT violation is developed.
We classify all Lorentz- and CPT-violating and invariant terms
in the quadratic Lagrange density for a Dirac fermion,
including operators of arbitrary mass dimension.
The exact dispersion relation is obtained in closed and compact form,
and projection operators for the spinors are derived.
The Pauli hamiltonians for particles and antiparticles are extracted,
and observable combinations of operators are identified.
We characterize and enumerate the coefficients for Lorentz violation 
for any operator mass dimension
via a decomposition using spin-weighted spherical harmonics.
The restriction of the general theory
to various special cases is presented, including 
isotropic models,
the nonrelativistic and ultrarelativistic limits,
and the minimal Standard-Model Extension.
Expressions are derived in several limits 
for the fermion dispersion relation,
the associated fermion group velocity,
and the fermion spin-precession frequency.
We connect the analysis to some other formalisms
and use the results to extract constraints 
from astrophysical observations
on isotropic ultrarelativistic spherical coefficients
for Lorentz violation.

\end{abstract}

\maketitle

\section{Introduction}

The invariance of the laws of nature under Lorentz transformations
is well established,
being based on an extensive series of investigations 
originating in classic tests 
such as the Michelson-Morley, Kennedy-Thorndike, Ives-Stilwell,
and Hughes-Drever experiments
\cite{mm,kt,is,hd}.
Interest in precision tests of relativity
has experienced a renewal in recent years,
following the realization that tiny departures 
from Lorentz invariance could arise 
in a fundamental theory such as strings
\cite{ksp}.
During this period,
experiments using techniques from many subfields 
have achieved striking sensitivities 
to a variety of effects from Lorentz violation
\cite{tables}.

The general framework characterizing violations of Lorentz invariance
is the Standard-Model Extension (SME)
\cite{ck,akgrav},
which is a realistic effective quantum field theory
incorporating General Relativity and the Standard Model.
Terms in the SME violating CPT symmetry also violate Lorentz invariance
\cite{owg},
so the SME also characterizes CPT violation.
Each Lorentz-violating term in the SME action
is a coordinate-independent scalar density
involving a Lorentz-violating operator
contracted with a controlling coefficient.
The mass dimension $d$ of the operator 
fixes the dimensionality of the corresponding coefficient.
In the popular scenario with General Relativity and the Standard Model 
emerging as the low-energy limit 
of an underlying theory of quantum gravity at the Planck scale
$M_P \sim 10^{19}$ GeV,
terms with larger $d$ can plausibly be viewed
as higher-order corrections
in a series approximating the underlying physics.
Other scenarios can also be envisaged. 

The focus of the present work is Lorentz violation in fermions.
The realistic nature of the SME means that it can readily be applied
to analyze observational and experimental data,
but existing studies of Lorentz violation with fermions 
are primarily concerned with the minimal SME,
obtained by restricting attention to operators 
of renormalizable dimensions $d\leq 4$.
To date,
the minimal SME has been adopted as the theoretical framework 
in searches for Lorentz violation in the fermion sector involving
electrons \cite{eexpt},
protons and neutrons \cite{pnexpt},
muons \cite{muexpt},
neutrinos \cite{nuexpt},
quarks
\cite{hadronexpt},
and gravitational couplings of various species 
\cite{akgrav,gravexpt}.
Discussions in the literature 
of the nonminimal SME fermion sector are more limited.
The general structure and properties of the nonminimal neutrino sector 
have been investigated
\cite{km12},
and results are known
for some special nonminimal SME-based models
\cite{rmmp,dm,rss,pbmp,ba11},
including ones with nonminimal fermion interactions
\cite{nonminfermionint}. 
However,
a complete description of the nonminimal SME fermion sector
remains an open issue.

In the present work,
we seek to address this gap in the literature
by extending the existing treatment of nonminimal Lorentz violation 
to include quadratic fermion operators of arbitrary mass dimension $d$,
thereby opening the path for additional searches for Lorentz violation.
To achieve a reasonable scope,
we restrict the analysis to flat spacetime
with a Dirac-type action invariant under spacetime translations
and phase rotations,
so that energy, momentum, and charge are conserved.
This scope suffices for applications
to many experimental situations involving fermions 
and can be applied to studies of matter
following methods used in the minimal sector
\cite{kla}.
It also serves as a basis for further theoretical investigations
of foundational aspects of Lorentz violation,
including mathematical topics 
such as the underling pseudo-Riemann-Finsler geometry 
\cite{finsler}
and physical issues such as causality and stability 
\cite{kle,causality},
where operators of large $d$ can dominate the associated physics.
Our results are also potentially relevant for some proposed theories
naturally generating effective field theories
dominated by SME operators of dimension $d>4$,
such as 
supersymmetric Lorentz-violating models 
\cite{susy}
or noncommutative quantum electrodynamics
\cite{hayakawa},
in which the corresponding SME operators have $d\geq 6$
\cite{chklo}.

The primary goal of this work is to develop
the quadratic nonminimal SME fermion sector
to the point where practical applications become feasible.
This requires extracting key information
from the general SME action,
including basic features of fermion behavior
in the presence of Lorentz violation.
Typical applications are expected to involve
measurements of aspects of fermion propagation,
such as times of flight or spin-precession rates,
and studies of fermion energy levels in systems such as atoms. 
The former require 
characterizing the anisotropy, dispersion, and birefringence 
in fermion propagation,
which can conveniently be addressed via the dispersion relation,
while the latter can be addressed 
by studying induced level shifts 
using the perturbative hamiltonian for Lorentz violation.
Here,
we obtain the exact dispersion relation and the perturbative hamiltonian,
and we develop a methodology to study the corresponding effects 
using a decomposition in spherical harmonics.
This permits a classification of all observables
in terms of four sets of coefficients for Lorentz violation
having straightforward rotation properties,
which is expected to simplify future experimental analyses.

This paper is organized as follows.
The general quadratic action for a Dirac field 
is studied in Sec.\ \ref{Single Dirac fermion}.
The basic framework is reviewed
in Sec.\ \ref{Basics},
while the role of field redefinitions
in determining physical observables is determined
in Sec.\ \ref{Field redefinitions}.
The exact vacuum dispersion relation 
is obtained in a closed and compact form
in Sec.\ \ref{Exact vacuum dispersion relation},
and some of its physical properties are described 
in Sec.\ \ref{Properties}.
Covariant projection operators for the spinor
solutions to the modified Dirac equation are presented 
in Sec.\ \ref{Spinors}.
We then turn to the construction of the hamiltonians
for particles and antiparticles,
deriving expressions for both
in Sec.\ \ref{Construction}
and converting them to explicitly covariant forms
in Sec.\ \ref{Coefficients}.
Taking advantage of the approximate rotational symmetry
relevant for many applications,
we perform in Sec.\ \ref{Spherical decomposition}
a decomposition of the hamiltonian in spin-weighted spherical harmonics.
This calculation yields a complete set 
of observable coefficients for Lorentz violation,
catalogued according to properties of the corresponding operators.
We develop the isotropic limit
for the perturbative hamiltonian
and present the general isotropic Lagrange density
for operator dimensions $d=3,4,5,6$
in both cartesian and spherical coefficients. 
In Sec.\ \ref{Limiting cases},
we turn to a description of various special cases of the framework,
including the nonrelativistic and ultrarelativistic limits
and the minimal SME.
Section \ref{Applications}
contains applications of the results
to dispersion, group velocity, and birefringence,
along with a discussion of connections between 
the nonminimal fermion sector of the SME
and other field theoretic and kinematical results
in the literature.
We also provide a compilation of existing astrophysical limits 
on isotropic Lorentz violation
translated into constraints 
on spherical SME coefficients.
Section \ref{Summary} summarizes the results obtained
in this work.
Throughout this paper,
we adopt conventions matching those 
of the prior studies of nonminimal Lorentz violation
in Refs.\ \cite{km09,km12}.

\section{Single Dirac fermion}
\label{Single Dirac fermion}

In this section, we consider the
effective action for a single Dirac fermion,
allowing for operators of arbitrary dimension.
Attention is restricted to terms 
that are quadratic in the fermion field,
which gives rise to a linear theory.
Features of the corresponding modified Dirac equation
are also considered,
including observability and field redefinitions. 
We derive an explicit expression 
for the exact dispersion relation,
and we determine an approximation
valid to leading order in Lorentz violation.
The result reveals features including
anisotropy, dispersion, and birefringence.
Leading-order expressions
for the eigenspinors are also presented.

\subsection{Basics}
\label{Basics}

Given the conventional Dirac Lagrange density,
the effective theory describing the fermion behavior
in the presence of general Lorentz violation
can be obtained by adding terms formed from tensor operators
contracted with coefficients for Lorentz violation
\cite{ck}.
The coefficients play the role of background fields 
generating the Lorentz violation,
and the resulting theory is coordinate independent.
For a single Dirac fermion $\ps$ of mass $\m$,
this construction and the requirement of a linear theory
imply that the action $S$ extends the usual Dirac action for $\ps$
by a quadratic functional of $\ps$ and its derivatives,
\bea
S &=& \int \cl ~d^4x ,
\nn\\
\cl &=& 
\half \psb (\ga^\mu i\prt_\mu - \m + \Qhat) \ps 
+ {\rm h.c.}, 
\label{lag}
\eea
where $\Qhat$ is a $4\times 4$ spinor-matrix operator
involving derivatives $i\prt_\mu$.
Without loss of generality, 
$\Qhat$ can be taken to obey the hermiticity condition
$\Qhat = \ga_0 \Qhat^\dag\ga_0$.
Since $\Qhat$ is general,
it includes both all Lorentz-invariant and all Lorentz-violating effects.
The latter may be Planck suppressed 
and in any case are generically tiny,
so we treat $\Qhat$ as a perturbative contribution 
when needed to insure that deviations 
from the conventional Dirac situation are small.
In particular,
this implies that any extra modes associated 
with the higher-order derivatives in $\Qhat$ can be neglected
for practical purposes.

The operator $\Qhat$ can describe effects of Lorentz violation 
arising either spontaneously or explicitly.
Spontaneous Lorentz violation occurs
when tensor fields dynamically acquire vacuum expectation values
\cite{ksp},
which play the role of background tensors in the operator $\Qhat$.
In contrast,
explicit Lorentz violation involves background tensors in $\Qhat$
that are externally prescribed.
In both cases,
the operator $\Qhat$ can in principle depend on spacetime position.
However,
to maintain invariance of the action \rf{lag}
under spacetime translations
and hence preserve energy-momentum conservation,
we require here that the operator $\Qhat$ is spacetime constant.
This insures a focus on pure Lorentz violation
and minimizes complications in analyses
at both the theoretical and experimental levels.
In the context of spontaneous Lorentz violation,
imposing spacetime translation symmetry
is equivalent to disregarding solitonic background fields
along with any massive and Nambu-Goldstone modes
\cite{ng}.
In certain situations these modes play the role of
the photon in Einstein-Maxwell theory 
\cite{akgrav,bumblebee},
the graviton
\cite{cardinal},
or other force mediators 
\cite{newforces,akjt},
so in these cases some care may be needed 
in interpreting results for spacetime-constant $\Qhat$. 

Note that spacetime constancy of $\Qhat$
can be either an exact feature of the model 
or an approximation to dominant or averaged effects
in a more complete theory.
The complete theory may even be fully Lorentz invariant.
Existing or hypothetical forces typically give rise to
effects with dominant contributions appearing as backgrounds
in a given experimental situation,
which can serve as effective Lorentz violation
in a phenomenological description.
For instance,
in a local laboratory the gravitational force produces 
a direction dependence that plays the role of explicit Lorentz violation
in the corresponding effective theory.
Hypothetical ultraweak forces can in principle be constrained
or even detected in this way.
For example,
sharp constraints on torsion have been obtained
by studying the effective Lorentz violation
associated with a torsion background 
\cite{torsion}.
In general,
viable models for Lorentz-invariant interactions
generating effective operators of the form $\Qhat$
must be consistent with known constraints on Lorentz violation
\cite{tables}.

In this subsection,
we perform a decomposition of $\Qhat$ 
that ultimately permits the enumeration and characterization
of the coefficients for Lorentz violation appearing
in the Lagrange density \rf{lag}.
Expanding $\Qhat$ in the basis of 16 Dirac matrices
explicitly reveals the spin content,
\bea
\Qhat &=& 
\sum_I \Qhat^I \ga_I
\nn\\
&=&
\Shat
+i\Phat \ga_5
+\Vhat^\mu \ga_\mu
+\Ahat^\mu \ga_5\ga_\mu
+\half \That^\mn \si_\mn ,
\label{qhatsplit}
\eea
where the 16 operators 
$\Qhat^I = \{\Shat,\Phat,\Vhat^\mu,\Ahat^\mu,\That^\mn\}$
are Dirac-scalar functions of the derivatives $i\prt_\mu$ 
with mass dimension one.
In momentum space,
each operator $\Qhat^I$ can be viewed 
as a series of terms,
\beq
\Qhat^I= \sum_{d=3}^\infty
\Q^{(d)I\al_1\al_2\ldots\al_{d-3}}
p_{\al_1}p_{\al_2}\ldots p_{\al_{d-3}} ,
\label{QI_expansion}
\eeq
with $p_\mu = i\prt_\mu$.
All the coefficients
$\Q^{(d)I\al_1\al_2\ldots\al_{d-3}}$
are spacetime independent and have dimension $4-d$.
Also,
they can all be assumed real by hermiticity.
Note that any of these coefficients proportional to combinations of products 
of the Lorentz-invariant tensors $\et^\mn$ and $\ep^{\ka\la\mn}$
correspond to Lorentz-invariant operators in the theory \rf{lag}.

\begin{table*}
\tabcolsep 5pt
\renewcommand{\arraystretch}{1.5}
\begin{tabular*}{0.8\textwidth}{*{6}{@{\extracolsep{\fill}}c}}
Operator	&	Type	&	$d$	&	CPT	&	Cartesian coefficients	&		Number		\\	\hline	\hline
$\mhat$	&	scalar	&	odd, $\geq 5$ 	&	even	&	$m^{(d)\al_1\ldots\al_{d-3}}$	&	$	d(d-1)(d-2)/6	$	\\	
$\mfivehat$	&	pseudoscalar	&	odd, $\geq 5$ 	&	even	&	$m_5^{(d)\al_1\ldots\al_{d-3}}$	&	$	d(d-1)(d-2)/6	$	\\	
$\ahat^\mu$	&	vector	&	odd, $\geq 3$ 	&	odd	&	$a^{(d)\mu\al_1\ldots\al_{d-3}}$	&	$	2d(d-1)(d-2)/3	$	\\	
$\bhat^\mu$	&	pseudovector	&	odd, $\geq 3$ 	&	odd	&	$b^{(d)\mu\al_1\ldots\al_{d-3}}$	&	$	2d(d-1)(d-2)/3	$	\\	
$\chat^\mu$	&	vector	&	even, $\geq 4$	&	even	&	$c^{(d)\mu\al_1\ldots\al_{d-3}}$	&	$	2d(d-1)(d-2)/3	$	\\		
$\dhat^\mu$	&	pseudovector	&	even, $\geq 4$	&	even	&	$d^{(d)\mu\al_1\ldots\al_{d-3}}$	&	$	2d(d-1)(d-2)/3	$	\\		
$\ehat$	&	scalar	&	even, $\geq 4$	&	odd	&	$e^{(d)\al_1\ldots\al_{d-3}}$	&	$	d(d-1)(d-2)/6	$	\\		
$\fhat$	&	pseudoscalar	&	even, $\geq 4$	&	odd	&	$f^{(d)\al_1\ldots\al_{d-3}}$	&	$	d(d-1)(d-2)/6	$	\\		
$\ghat^\mn$	&	tensor	&	even, $\geq 4$	&	odd	&	$g^{(d)\mn\al_1\ldots\al_{d-3}}$	&	$	d(d-1)(d-2)	$	\\		
$\Hhat^\mn$	&	tensor	&	odd, $\geq 3$ 	&	even	&	$H^{(d)\mn\al_1\ldots\al_{d-3}}$	&	$	d(d-1)(d-2)	$	\\	\hline	
\end{tabular*}
\caption{\label{free_summary}
Operators and coefficients for a Dirac fermion.}
\end{table*}

Often,
it is convenient to work with an alternative decomposition of $\Qhat$
that parallels the formalism widely used 
for the single-fermion limit of the minimal SME
\cite{ck}.
This parallel suggests writing 
\beq
\ga^\nu p_\nu - \m + \Qhat 
= \widehat\Ga^\nu p_\nu - \widehat M ,
\label{qhat}
\eeq
where $\widehat\Ga^\nu p_\nu$ and $\widehat M$ 
consist of operators of even and odd mass dimension,
respectively.
Decomposing these operators in terms of the basis
of 16 Dirac matrices yields
\bea
\widehat\Ga^\nu &=& 
\ga^\nu
+\chat^{\mn} \ga_\mu
+\dhat^{\mn} \ga_5\ga_\mu
+\ehat^\nu
+i\fhat^\nu \ga_5
+\half \ghat^{\ka\la\nu} \si_{\ka\la} ,
\nn\\
\widehat M &=& 
\m 
+ \mhat
+i\mfivehat \ga_5
+\ahat^\mu \ga_\mu
+\bhat^\mu \ga_5\ga_\mu
+\half \Hhat^\mn \si_\mn .
\nn\\
\label{GaM}
\eea
In these expressions,
the operators
$\chat^{\mn}$, $\dhat^{\mn}$
are CPT even and dimensionless,
$\ehat^\mu$, $\fhat^\mu$, $\ghat^{\mu\rh\nu}$
are CPT odd and dimensionless,
$\mhat$, $\mfivehat$, $\Hhat^\mn$ 
are CPT even and of dimension one, 
and $\ahat^\mu$, $\bhat^\mu$
are CPT odd and of dimension one.
If desired,
a chiral mass term $i m_5 \ga_5$ can be added to $\widehat M$,
but in many situations 
this can be absorbed into $\m$ via a chiral rotation
without loss of generality
and so we omit it from Eq.\ \rf{GaM}.
The operators $\mhat$ and $\mfivehat$
consist solely of higher-derivative terms of nonrenormalizable dimension,
but all the others appearing in Eq.\ \rf{GaM}
have equivalents in the minimal SME.

In Eq.\ \rf{qhat},
the operator $\widehat\Ga^\nu$ is contracted with $p_\nu$.
This implies that the operators 
$\chat^{\mn}$, $\dhat^{\mn}$,
$\ehat^\mu$, $\fhat^\mu$, $\ghat^{\mu\rh\nu}$
are also contracted with $p_\nu$,
and it motivates the introduction of contracted operators via 
\bea
\chat^\mu &=& \chat^{\mn} p_\nu  , 
\quad
\dhat^\mu = \dhat^{\mn} p_\nu ,
\nn\\
\ehat &=& \ehat^\nu p_\nu ,
\quad
\fhat = \fhat^\nu p_\nu , 
\quad
\ghat^{\ka\la} = \ghat^{\ka\la\nu} p_\nu .
\eea
The notation for each operator has been chosen 
so that its CPT handedness corresponds to that
of its analogue in the minimal SME. 
In terms of these operators,
we find
\bea
\Shat &=& \ehat - \mhat ,
\quad
\Phat = \fhat - \mfivehat,
\quad
\Vhat^\mu = \chat^\mu - \ahat^\mu , 
\nn\\ 
\Ahat^\mu &=& \dhat^\mu - \bhat^\mu , 
\quad
\That^\mn = \ghat^\mn - \Hhat^\mn.
\label{rels}
\eea
These expressions provide the explicit link
between the decompositions \rf{qhatsplit} and \rf{qhat}.

Each of the 10 component operators
$\ehat$, 
$\mhat$, 
$\fhat$, 
$\mfivehat$, 
$\chat^{\mu}$, 
$\ahat^\mu$, 
$\dhat^{\mu}$, 
$\bhat^\mu$, 
$\ghat^{\mn}$,
$\Hhat^\mn$
can be expanded in cartesian momentum components
following the form of Eq.\ \rf{QI_expansion},
yielding ten infinite series of real coefficients.
For example,
the operator $\chat^\mu$ can be written as
\beq
\chat^\mu = 
\sum_{d {\rm ~even}} 
c^{(d)\mu\al_1\ldots\al_{d-3}} p_{\al_1}\ldots p_{\al_{d-3}} .
\eeq
Each term in this sum involves a coefficient
$c^{(d)\mu\al_1\ldots\al_{d-3}}$,
for which the index $\mu$ controls the spin nature of the operator
and the $d-3$ symmetric indices $\al_1\ldots\al_{d-3}$
control the momentum dependence.
In the analogous expansions for the photon sector
\cite{km09},
the spin and momentum dependence 
are intertwined by gauge symmetry,
which complicates the counting of components.
Here,
however,
the number of independent coefficients 
in $c^{(d)\mu\al_1\ldots\al_{d-3}}$ for each $d$ 
can be obtained directly as $2 d(d-1)(d-2)/3$.
The coefficients for the nine other operators
can be treated similarly.
Table \ref{free_summary}
lists the 10 operators, 
their corresponding coefficients,
and some of their properties.

\subsection{Field redefinitions}
\label{Field redefinitions}

In the context of the minimal SME,
the freedom to choose coordinates and to redefine fields 
while leaving the physics unchanged 
makes some coefficients for Lorentz violation physically unobservable 
\cite{ck,akgrav,akjt,km02,redefref,ba-f,classical}.
This feature extends to the nonminimal sector.
The effects of a coordinate choice,
which amounts to selecting the sector in which 
the effective background spacetime metric
has the usual diagonal Minkowski form,
are analogous to those in the minimal SME
and imply 10 combinations of coefficients are always unobservable. 
Also,
the freedom to redefine the fermion $\ps$ by a position-dependent phase,
\beq
\ps = \exp(ix^\mu v_\mu)\ps'
\label{phaseredef}
\eeq 
for a suitable $v_\mu$,
can be used as in the minimal SME to remove four constant coefficients
coupling like a gauge potential.
However,
the freedom to make field redefinitions involving the spinor space,
which eliminates and recombines certain coefficients in the Lagrange density,
is more involved when the nonminimal sector is incorporated.

Here,
we consider field redefinitions of the form
\beq
\ps = (1+\Zhat) \ps' ,
\label{redef}
\eeq
where $\Zhat$ is an arbitrary $p$-dependent operator.
For this redefinition to leave the physics unaffected,
the dominant modes in the Lagrange density
must remain dominant in the redefined theory, 
and so the perturbative assumption for the operator $\Qhat$ 
in the Dirac action \rf{lag} must be maintained.
This implies that $\Zhat$ itself must be perturbative.

Under the redefinition \rf{redef},
the operator in the Dirac action \rf{lag}
acquires a new form, 
\beq
\ps^\dag \ga_0(p\cdot\ga - \m + \Qhat) \ps
\approx
\ps'^\dag \ga_0(p\cdot\ga - \m + \Qhat') \ps' ,
\eeq
where
\beq
\Qhat' = \Qhat 
+ (p\cdot\ga - \m) \Zhat + \ga_0\Zhat^\dag\ga_0 (p\cdot\ga - \m) .
\eeq
To explore the implications of this structure,
it is useful to split $\Zhat$ 
into a hermitian piece $\Xhat$ and an antihermitian piece $\Yhat$,
defined according to 
\beq
\Zhat = \Xhat+i\Yhat , 
\hskip 4pt
\Xhat = \half(\Zhat+\ga_0\Zhat^\dag\ga_0) , 
\hskip 4pt
\Yhat = \tfrac{1}{2i}(\Zhat-\ga_0\Zhat^\dag\ga_0) ,
\eeq
where both $\Xhat$ and $\Yhat$ obey the same hermiticity
condition as the $\Qhat$ operator,
$\Xhat = \ga_0 \Xhat^\dag\ga_0$,
$\Yhat = \ga_0 \Yhat^\dag\ga_0$.
The operator $\Qhat'$ is then given by
\beq
\Qhat' = \Qhat -2\m\Xhat + p_\mu \{\ga^\mu,\Xhat\} +ip_\mu[\ga^\mu,\Yhat] .
\eeq
This shows that a suitable choice of $\Xhat$ or $\Yhat$
can combine with $\Qhat$
to reduce the observable content of $\Qhat'$.
 
To determine explicitly which pieces of $\Qhat$ are affected,
we can decompose both $\Xhat$ and $\Yhat$ 
in the basis of 16 Dirac matrices,
\bea
\Xhat &=& \Xhat_S
+ i\Xhat_P\ga_5
+ \Xhat_V^\mu \ga_\mu
+ \Xhat_A^\mu \ga_5\ga_\mu
+ \half \Xhat_T^\mn \si_\mn ,
\nn\\
\Yhat &=& \Yhat_S
+ i\Yhat_P \ga_5
+ \Yhat_V^\mu \ga_\mu
+ \Yhat_A^\mu \ga_5\ga_\mu
+ \half \Yhat_T^\mn \si_\mn .
\qquad
\eea
Each component of $\Xhat$ and $\Yhat$ in these expansions
can be considered independently.
The component $\Yhat_S$ evidently has no effect on $\Qhat$,
but the other nine generate field redefinitions 
mixing various Lorentz-violating operators
and acting to remove some of them at leading order.
In what follows,
we apply each field redefinition in turn,
determining the changes $\de\Qhat = \Qhat'-\Qhat$
and identifying the resulting effects.

First,
consider transformations involving the components of $\Xhat$. 
A nonzero $\Xhat_S$ gives
\beq
\de\Shat = -2\m\Xhat_S , 
\quad
\de\Vhat^\mu = 2\Xhat_S p^\mu ,
\eeq
showing that for $\m\neq 0$ the scalar operator $\Shat$ 
can be removed by absorbing it into the vector $\Vhat^\mu$.
Using $\Xhat_P$ instead produces 
\beq
\de\Phat = -2\m\Xhat_P ,
\eeq
which reveals that the pseudoscalar operator $\Phat$ can also be removed.
A nonzero $\Xhat_V^\mu$ gives
\beq
\de\Shat = 2p_\mu\Xhat_V^\mu , 
\quad
\de\Vhat^\mu = -2\m\Xhat_V^\mu ,
\eeq
which reconfirms that the scalar and vector operators 
mix under field redefinitions.
Using $\Xhat_A^\mu$ yields 
\beq
\de\Ahat^\mu = -2\m\Xhat_A^\mu , 
\quad
\de\That^\mn = -2\ep^{\mn\rh\si}p_\rh\Xhat_{A\si} ,
\eeq
so the pseudovector operators $\Ahat^\mu$ 
can be absorbed into the tensor ones.
Finally,
a nonzero $\Xhat_T^\mn$ gives
\beq
\de\Ahat^\mu = -\ep^{\mn\rh\si}p_\nu \Xhat_{T\rh\si} , 
\quad
\de\That^\mn = -2\m\Xhat_T^\mn ,
\eeq
again showing that the pseudovector and tensor operators mix.

Next,
we turn to transformations involving the components of $\Yhat$. 
Taking nonzero $\Yhat_S$ has no effect,
as mentioned above.
Using $\Yhat_P$ gives 
\beq
\de\Ahat^\mu = 2p^\mu\Yhat_P ,
\eeq
which permits the removal 
of the component of $\Ahat^\mu$ 
proportional to $p^\mu$.
A nonzero $\Yhat_V^\mu$ gives 
\beq
\de\That^\mn = 2 p^{[\mu} \Yhat_V^{\nu]} .
\eeq
In the minimal sector,
this can be used to remove the trace component of $g^{(4)\mn\rh}$.
More generally, 
the coefficients appearing in the expansion of the dual 
$\Tdual^\mn = \half\ep^{\mn\rh\si} \That_{\rh\si}$
can be split into pieces that transform under
two different representations of the Lorentz group,
with one set antisymmetric in the first three indices
and the other antisymmetric in the first two indices
with vanishing antisymmetrization on any three indices.
The above field redefinition with $\Yhat_V^\mu$ 
can be used to remove the first piece.
Taking instead a nonzero $\Yhat_A^\mu\neq 0$ gives
\beq
\de\Phat = -2p_\mu Y_A^\mu ,
\eeq
which reconfirms that the pseudoscalar operator $\Phat$ can be removed.
Finally, 
using $\Yhat_T^\mn$ leads to
\beq
\de\Vhat^\mu = 2\Yhat_T^\mn p_\nu .
\eeq
In this case,
the coefficients appearing in the expansion of $\Vhat^\mu$ 
can be split into a piece that is totally symmetric
and one with mixed symmetry that is antisymmetric in the first two indices.
The field redefinition with $\Yhat_T^\mn$ 
allows the removal of the piece with mixed symmetry.

We thus see that the physical observables 
in the quadratic fermion theory \rf{lag}
are restricted to pieces of $\Vhat^\mu$ and $\That^\mn$.
The relationships \rf{rels} show
these observables correspond to parts of 
$\ahat^\mu$, $\chat^\mu$, $\ghat^\mn$, and $\Hhat^\mn$.
This feature parallels results for the neutrino sector,
where the propagation of neutrinos is controlled by
four effective coefficients of these types 
despite the multiple flavors, the mixing, and
the handedness of the fermions
\cite{km12}.
It also reduces correctly to known results in the minimal SME
\cite{ck,akgrav}.

\begin{table*}	
\tabcolsep 5pt
\renewcommand{\arraystretch}{1.5}
\begin{tabular*}{0.8\textwidth}{*{6}{@{\extracolsep{\fill}}c}}
Operator	  &	Type	&	$d$	&	CPT	&	Cartesian coefficients	&		Number		\\	\hline	\hline
$\ahat_\eff^\mu$  &	vector	&	odd, $\geq 3$	&	odd	&	$a_\eff^{(d)\mu\al_1\ldots\al_{d-3}}$	&	$(d+1)d(d-1)/6$	\\
$\chat_\eff^\mu$  &	vector	&	even, $\geq 4$	&	even	&	$c_\eff^{(d)\mu\al_1\ldots\al_{d-3}}$	&	$(d+1)d(d-1)/6$	\\
$\gdual_\eff^\mn$ &	tensor	&	even, $\geq 4$	&	odd	&	$\gt_\eff^{(d)\mn\al_1\ldots\al_{d-3}}$	&	$(d+1)d(d-2)/2$	\\		
$\Hdual_\eff^\mn$ &	tensor	&	odd, $\geq 3$ 	&	even	&	$\Ht_\eff^{(d)\mn\al_1\ldots\al_{d-3}}$	&	$(d+1)d(d-2)/2$ \\	\hline	
\end{tabular*}
\caption{\label{effecive_summary}
Effective operators and effective coefficients for a Dirac fermion.}
\end{table*}

Using the field redefinitions,
we can define a canonical set of effective operators
representing physical observables in the quadratic theory \rf{lag},
\bea
\Shat_\eff &=& 0, 
\quad 
\Phat_\eff = 0 , 
\quad 
\Ahat_\eff^\mu = 0 ,
\nn\\
\Vhat_\eff^\mu &=& 
\big(\Vhat^\mu + \tfrac{1}{\m} p^\mu \Shat \big)_{[0]} ,
\nn\\
\Tdual_\eff^\mn &=& 
\big(\Tdual^\mn + \tfrac{1}{\m} p^{[u}\Ahat^{\nu]}\big)_{[2]} ,
\label{effops}
\eea
where the subscript $[n]$ indicates 
that the coefficients appearing in the operator expansion
are restricted to an irreducible representation
antisymmetric in the first $n$ indices.
The relationships \rf{rels} imply the corresponding definitions 
\bea
\ahat_\eff^\mu &=& 
\big( \ahat^\mu - \tfrac{1}{\m} p^\mu \ehat \big)_{[0]}
\nn\\
&=& \sum_d a_\eff^{(d)\mu\al_1\ldots\al_{d-3}}p_{\al_1}\ldots p_{\al_{d-3}} ,
\nn\\
\chat_\eff^\mu
&=& (\chat^\mu - \tfrac{1}{\m} p^\mu \mhat)_{[0]}
\nn\\
&=& \sum_d c_\eff^{(d)\mu\al_1\ldots\al_{d-3}}p_{\al_1}\ldots p_{\al_{d-3}},
\nn\\
\gdual_\eff^\mn
&=& (\gdual^\mn - \tfrac{1}{\m} p^{[\mu} \bhat^{\nu]} \big)_{[2]}
\nn\\
&=& \sum_d \gt_\eff^{(d)\mn\al_1\ldots\al_{d-3}}p_{\al_1}\ldots p_{\al_{d-3}},
\nn\\
\Hdual_\eff^\mn
&=& (\Hdual^\mn - \tfrac{1}{\m} p^{[\mu} \dhat^{\nu]} \big)_{[2]}
\nn\\
&=& \sum_d \Ht_\eff^{(d)\mn\al_1\ldots\al_{d-3}}p_{\al_1}\ldots p_{\al_{d-3}} .
\label{effcompops}
\eea
Note that the analysis of field redefinitions 
naturally leads to expansions of the duals
$\gdual_\eff^\mn$, $\Hdual_\eff^\mn$
rather than the tensor operators 
$\ghat_\eff^\mn$, $\Hhat_\eff^\mn$ directly.

In terms of the fundamental coefficients,
the effective coefficients are
\bea
a_\eff^{(d)\mu\al_1\ldots\al_{d-3}}
  &=& \big(a^{(d)\mu\al_1\ldots\al_{d-3}}
\nn\\
&&
\hskip 30pt
- \tfrac{1}{\m} \et^{\mu\al_1}e^{(d-1)\al_2 \ldots \al_{d-3}}\big)_{[0]},
\nn\\
c_\eff^{(d)\mu\al_1\ldots\al_{d-3}}
&=& \big(c^{(d)\mu\al_1\ldots\al_{d-3}}
\nn\\
&&
\hskip 30pt
- \tfrac{1}{\m} \et^{\mu\al_1}m^{(d-1)\al_2 \ldots \al_{d-3}}\big)_{[0]},
\nn\\
\gt_\eff^{(d)\mn\al_1\ldots\al_{d-3}}
&=& \big(\gt^{(d)\mn\al_1\ldots\al_{d-3}}
\nn\\
&&
\hskip 30pt
- \tfrac{1}{\m} \et^{\al_1[\mu}b^{(d-1)\nu]\al_2 \ldots \al_{d-3}}\big)_{[2]},
\nn\\
\Ht_\eff^{(d)\mn\al_1\ldots\al_{d-3}}
&=& \big(\Ht^{(d)\mn\al_1\ldots\al_{d-3}}
\nn\\
&&
\hskip 30pt
- \tfrac{1}{\m} \et^{\al_1[\mu}d^{(d-1)\nu]\al_2 \ldots \al_{d-3}}\big)_{[2]}.
\nn\\
\label{effcoeffcomps}
\eea
In these equations,
the dual coefficients are defined by 
\bea
\gt^{(d)\mn\al_1\ldots\al_{d-3}}
&=& \half {\ep^\mn}_{\rh\si} g^{(d)\rh\si\al_1\ldots\al_{d-3}} , 
\nn\\
\Ht^{(d)\mn\al_1\ldots\al_{d-3}}
&=& \half {\ep^\mn}_{\rh\si} H^{(d)\rh\si\al_1\ldots\al_{d-3}} .
\eea
Also,
the subscript $[0]$ indicates symmetrization on all indices,
while $[2]$ indicates symmetrization on $\nu\al_1\ldots\al_{d-3}$
followed by antisymmetrization on $\mn$.

The above results demonstrate,
for example, 
that leading-order signals from $\bhat^\mu$ 
can be absorbed into those from $\gdual^\mn$, 
while signals from $\dhat^\mu$ 
merge with those of $\Hdual^\mn$.
As an illustration, the $d=4$ terms in $\dhat^\mu$ 
can be absorbed into the $d=5$ terms in $\Hdual^\mn$,
giving rise to effective coefficients
$\Ht_\eff^{(5)\mu\al_1\al_2\al_3}$.
This example also reveals the potentially surprising result
that an operator naively having renormalizable dimension
and hence lying in the minimal SME
may in fact most naturally be regarded
as belonging to the nonminimal sector 
and having nonrenormalizable dimension.
Note also that the cartesian coefficients
$m_5^{(d)\al_1\ldots\al_{d-3}}$ and
$f^{(d)\al_1\ldots\al_{d-3}}$
have no observable role.
This is consistent with known results for the minimal case
\cite{akgrav,ba-f,classical}.
Moreover,
additional field redefinitions or coordinate choices 
can further reduce the number of observable effects.
For example,
the phase redefinition \rf{phaseredef}
shows that the effective coefficient $a_\eff^{(3)\mu}$
is unobservable.
All the effective operators,
their cartesian effective coefficients, and some of their properties
are compiled in Table \ref{effecive_summary}.

A few of the effective coefficients correspond 
to Lorentz-invariant operators in the theory \rf{lag}.
They must be formed from 
combinations of products of the Lorentz-invariant tensors
$\et^\mn$ and $\ep^{\mn\rh\si}$
multiplied by constant scalars.
The coefficients
$a_\eff^{(d)\mu\al_1\ldots\al_{d-3}}$
and
$\gt_\eff^{(d)\mn\al_1\ldots\al_{d-3}}$
both have an odd number of indices,
so they all produce Lorentz-violating effects.
Inspection reveals that the symmetries of
the coefficients $\Ht_\eff^{(d)\mn\al_1\ldots\al_{d-3}}$
preclude constructing them in terms of invariant tensors as well.
The only option for generating Lorentz-invariant operators 
is therefore to use the coefficients
$c_\eff^{(d)\mu\al_1\ldots\al_{d-3}}$
constructed as completely symmetrized products of the metric,
\beq
c_{\eff,{\rm LI}}^{(d)\mu\al_1\ldots\al_{d-3}} = 
\fr{1}{(d-2)!} c^{(d)}_{\rm LI}
\et^{(\mu\al_1}\et^{\al_2\al_3}
\cdots\et^{\al_{d-4}\al_{d-3})} .
\label{lorinv}
\eeq
This reveals that there is exactly
one Lorentz-invariant effective operator 
at each even dimension $d=4,6,\ldots$.
No Lorentz-invariant effective operators
exist for odd $d$.

The addition of interactions 
or the presence of a non-Minkowski background
typically changes the set of physical observables
by affecting the implementation of field redefinitions.
For generality in what follows,
we therefore present calculations and results 
with all coefficients explicitly included.
However,
expressions relevant for physical measurements
can be expected to yield only observable quantities. 
For example,
the effective coefficients appearing in the hamiltonian 
derived in Sec.\ \ref{Hamiltonian} below 
are compatible with this structure of observables.

\subsection{Exact vacuum dispersion relation}
\label{Exact vacuum dispersion relation}

The action \rf{lag} leads to the modified Dirac equation  
\beq
(p\cdot\ga - \m + \Qhat) \ps = 0,
\label{moddirac}
\eeq
where $\Qhat$ can be viewed as the expression \rf{qhatsplit}.
Formally,
the exact dispersion relation 
for plane-wave solutions in the vacuum 
is found by requiring 
that the determinant of the modified Dirac operator vanishes,
\beq
\det(p\cdot\ga - \m + \Qhat) = 0.
\label{dr}
\eeq
This condition determines the propagation
of spinor wave packets 
in the presence of Lorentz-violating operators
of arbitrary dimension.

An explicit form for the dispersion relation \rf{dr}
can be obtained by direct calculation.
One method proceeds by adopting a chiral representation 
of the Dirac matrices
and breaking the modified Dirac operator
into $2\times 2$ blocks $A$, $B$, $C$, $D$,
\bea
p\cdot\ga - \m + \Qhat 
&=&
\begin{pmatrix} A & B \\ C & D \end{pmatrix} .
\label{block}
\eea
It is convenient to introduce the notation
$\si^\mu = (\si^0, \si^j)$,
where $\si^0$ is the $2\times 2$ identity matrix
and $\si^j$ are the usual three Pauli matrices.
The adjoint matrices are 
$\bsi^\mu = (\si^0, -\si^j)$,
and they satisfy the basic identity
\bea
\bsi_\mu\si_\nu &=& 
\et_\mn + \frac{i}{2}\ep_{\mu\nu\ka\la} \bsi^\ka\si^\la .
\label{pauli} 
\eea
The block decomposition can then be written
\beq
\begin{pmatrix} A & B \\ C & D \end{pmatrix}
=
\begin{pmatrix}
\Shat_- + \frac{i}{2}\That_-^\mn \si_\mu\bsi_\nu &
\Vhat_-^\mu \si_\mu \\
\noalign{\medskip}
\Vhat_+^\mu \bsi_\mu &
\Shat_+ + \frac{i}{2}\That_+^\mn \bsi_\mu\si_\nu \\
\end{pmatrix}, 
\quad
\eeq
where
\bea
\Shat_\pm &=& - \m + \Shat \pm i \Phat ,
\nn\\
\Vhat^\mu_\pm &=& p^\mu + \Vhat^\mu \pm  \Ahat^\mu ,
\nn\\
\That^\mn_\pm &=& \half(\That^\mn \pm i\Tdual^\mn).
\eea
Here,
$\Tdual^\mn$ is the dual of $\That^\mn$.
Note that $\That^\mn_\pm = \pm i \Tdual^\mn_\pm$
are the two chiral components
of the tensor operator $\That^\mn$.

The determinant \rf{dr} can be obtained
from the block form \rf{block}
using the identity
\beq
\det \begin{pmatrix} A & B \\ C & D \end{pmatrix}
= \det (AD) + \det (BC) - \tr(\ol BA\ol CD) ,
\label{blockdet}
\eeq
where
$\ol B = {\rm adj} (B) $ and $\ol C = {\rm adj} (C)$
are matrix adjoints.
This quantity can be directly evaluated
using the basic result \rf{pauli} 
and the subsidiary identities 
\bea
\tr(\bsi_\mu\si_\nu\bsi_\ka\si_\la) &=& 
2\et_{\mu\nu\ka\la} ,
\nn\\
\tr(\bsi_\mu\si_\nu\bsi_\ka\si_\la\bsi_\rh\si_\si)
&=& 
2{\et_{\mu\nu\ka}}^\ta \et_{\ta\la\rh\si} ,
\nn\\
\et_{\mu\nu\ka\la}T_-^{\nu\ka} = 4T_{-\mu\la} ,
&&
\et_{\mu\nu\ka\la}T_+^{\ka\la} = -4T_{+\mu\nu} ,
\label{miscidentities}
\eea
where $\et_{\mu\nu\ka\la}$ is defined by
\bea
\et_{\mu\nu\ka\la} &=&
\et_{\mu\nu}\et_{\ka\la}
-\et_{\mu\ka}\et_{\nu\la}
+\et_{\mu\la}\et_{\nu\ka}
-i\ep_{\mu\nu\ka\la} .
\eea

The calculation outlined above 
yields an explicit form for the exact dispersion relation \rf{dr}
of the modified Dirac operator.
We find
\bea
&
\hskip -100pt
(\Shat_-^2-\That_-^2)(\Shat_+^2-\That_+^2)
+\Vhat_-^2\Vhat_+^2
\nn\\
& 
-2\Vhat_- \cdot
\big(\Shat_-+2i\That_-\big) \cdot
\big(\Shat_+-2i\That_+\big) \cdot
\Vhat_+  = 0 ,
\label{exactdr}
\eea
where $\That_\pm^2 = \That_\pm^\mn\That_{\pm\mn}$.
This compact expression holds for a Dirac field
experiencing Lorentz violation
involving operators of arbitrary mass dimension.
It reduces correctly to the well-known result 
for the renormalizable theory 
\cite{kle}
and its nonrelativistic limit
\cite{kla}.
In terms of the effective operators \rf{effops},
we obtain
\beq
\hskip -50pt
0 = (p + \Vhat_\eff)^4 + (\m^2-\That_{\eff -}^2)(\m^2-\That_{\eff +}^2)
\nn
\eeq
\vskip -25pt
\beq
-2(p + \Vhat_\eff) \cdot \big(\m - 2 i\That_{\eff -}\big)
\cdot \big(\m + 2 i\That_{\eff +}\big) \cdot (p + \Vhat_\eff) ,
\eeq
where
$\That^\mn_{\eff\pm} = \half(\That_\eff^\mn \pm i\Tdual_\eff^\mn)$.

The superficially quartic nature of the dispersion relation 
\rf{exactdr}
reflects the usual presence of the four independent Dirac spinors,
representing two spin projections 
for each of the particle and antiparticle modes.
However,
viewed as a function of $p_\mu$,
the dispersion relation \rf{exactdr}
represents an algebraic variety $\cR(p_\mu)$ 
of arbitrarily high order
rather than the usual Dirac quartic.
When the coefficients for Lorentz violation are small,
four roots of $\cR$ appear as small corrections
to the four roots of the usual Dirac equation,
while the remaining roots represent high-frequency modes
that are physically uninteresting. 
This behavior is analogous to that 
found for the exact covariant dispersion relation
for photons in the presence 
of Lorentz-violating operators of arbitrary dimension,
which is given as Eq.\ (30) of Ref.\ \cite{km09}. 
We remark in passing that 
the explicit dispersion relation \rf{exactdr} 
can be expected to have an interpretation
in terms of the geodesic motion of a classical particle
in a Finsler spacetime,
paralleling the existing treatment of the renormalizable case 
\cite{finsler,classical}.

\subsection{Properties}
\label{Properties}

For many practical purposes,
and to gain insight about the physical content
of the exact result \rf{exactdr},
it is useful to consider 
the approximate dispersion relation
valid at leading order in Lorentz violation.
Since Lorentz violation is expected to be small,
$\Qhat$ can be taken as a perturbation on
the conventional Dirac operator 
$p\cdot\ga - \m$.
The dispersion relation can therefore be expanded 
in the small operators
$\Shat$,
$\Phat$,
$\Vhat^\mu$,
$\Ahat^\mu$,
$\That^\mn$.

At leading order,
this expansion yields the approximate dispersion relation 
\bea
p^2 - \m^2 &\approx &
2( - \m\Shat -p\cdot\Vhat \pm \BD) ,
\label{drapprox}
\eea
where 
\bea
\BD^2 &=& 
(p\cdot\Ahat)^2 
-\m^2 \Ahat^2
-2\m p\cdot\Tdual\cdot\Ahat
+p\cdot\Tdual\cdot\Tdual\cdot p 
\nn\\
&=&
p\cdot\Tdual_\eff\cdot\Tdual_\eff\cdot p .
\label{deqn}
\eea
Solving for the energy $\eom$ gives
\bea
\eom &\approx &
\eomz - \fr{ \m \Shat + p\cdot\Vhat }{\eomz} 
\pm \fr{\BD}{\eomz} 
\nn\\
&=&
\eomz - \fr{ p\cdot\Vhat_\eff }{\eomz} 
\pm \fr{\BD}{\eomz} ,
\label{drapprox2}
\eea
where
$\eomz^2 = \m^2 + \mbf p^2$.
The results in this section are valid for $\eomz$ of either sign,
but in subsequent sections we take $\eomz > 0$.
Note that the terms $\Vhat$, $\Shat$, and $\BD$
depend on the 4-momentum,
which at the relevant order in Lorentz violation 
can be taken as $p^\mu \approx (\eomz, \mbf p)$
on the right-hand side of this equation.

The two sign choices for $\eomz$ 
correspond to particle and antiparticle modes,
so in the presence of nonzero Lorentz violation
the dispersion relation \rf{drapprox2} 
can have four nondegenerate solutions for each $\mbf p$. 
The usual spin degeneracy of a free Dirac fermion
is broken when $\BD$ is nonzero,
which requires pseudovector or tensor operators 
for Lorentz violation.
In contrast,
the scalar and vector operators for Lorentz violation
can shift the energy but preserve the spin degeneracy.
The degeneracy between particles and antiparticles
is broken when any of these operators have
nonzero CPT-odd components.
Note,
however,
that the pseudoscalar operator plays no role
in the leading-order dispersion relation. 

The dispersion relation \rf{drapprox2}
describes various kinds of deviations 
from the conventional Lorentz-covariant behavior 
of a massive fermion.
Many are analogous to effects appearing 
in the nonminimal photon sector of the SME
\cite{km09,nonminphot}.
Among them are anisotropy,
dispersion,
and birefringence.

Anisotropy is a consequence of violation of rotation invariance,
which implies the properties of the fermion 
depend on the momentum orientation $\phat$.
For example, 
the group velocity 
$\mbf v_g = {\prt E}/{\prt\mbf p}$
becomes a direction-dependent quantity.
We emphasize that in practice anisotropy 
is always present in models with physical Lorentz violation,
even ones formulated as being rotation invariant 
in a particular frame,
because boosts induce rotations.
For example,
any laboratory frame is instantaneously boosted 
by the Earth's rotation and revolution about the Sun,
and these boosts necessarily introduce anisotropy. 

When a dispersion relation is nonlinear,
component waves in a packet travel 
at different phase velocities $\mbf v_p = \mbf p/E$.
This dispersion is a familiar feature 
for a conventional massive fermion,
and most Lorentz-violating operators are dispersive.
Indeed,
the only nondispersive terms in the Lagrange density \rf{lag}
are those with a single derivative.
However,
certain dispersive terms are unobservable
at leading order in Lorentz violation.
For example,
the dispersive operators
contained in the pseudoscalar $\Phat$ play no role 
in the dispersion relation \rf{drapprox2}.
Also,
the restriction $E\approx \eomz$
on the right-hand side of this equation
implies that some dispersive operators 
in $\Vhat$, $\Shat$, and $\BD$ 
produce effects in vacuum propagation 
that are unobservable at leading order.
Changing the boundary conditions or introducing a medium
leaves unaffected the basic dispersive nature of an operator,
which is associated with its derivative structure.
However,
the corresponding change in the physics 
can trigger dispersion controlled 
by coefficients for Lorentz violation 
that at leading order are unobservable in the vacuum case.
This is analogous to the situation for photon propagation
\cite{km09}.

In the presence of Lorentz violation,
the fermion spin projections can mix during propagation
because spin may no longer be conserved.
Following the terminology for the analogous mixing of photon spins 
in Lorentz-violating electrodynamics,
we refer to this spin mixing as fermion birefringence.
It occurs whenever 
a particular solution to the dispersion relation
is associated with only one low-energy mode
instead of the usual two degenerate spin modes.
The dispersion relation \rf{drapprox2}
holds for plane-wave solutions
obeying the usual boundary conditions for vacuum propagation,
and its form implies that fermion birefringence occurs 
whenever the combination $\BD$ of coefficients 
given in Eq.\ \rf{deqn} is nonzero.
Note,
however,
that other situations 
such as a fermion trapped in a spherical container
can involve different boundary conditions
and hence can lead to spin-mixing effects 
controlled by different combinations of coefficients.
Also,
the presence of a medium 
such as matter or a background electromagnetic field
can be expected to modify the combination of coefficients
controlling fermion birefringence,
which again parallels the situation 
for Lorentz-violating effects in photon propagation
\cite{km09}.

\subsection{Spinors}
\label{Spinors}

A basic feature of the conventional Dirac equation
is that the four linearly independent eigenspinors 
can be written using covariant projection operators as 
\beq
u_\pm (p,n) = P_\pm \La_+ \ps,
\quad
v_\pm (p,n) = P_\mp \La_- \ps,
\label{eigenspinors}
\eeq
where the projection operators $\La_\pm$ select 
positive- and negative-energy states,
while 
\beq
P_\pm = \half(1\pm \ga_5 n\cdot\ga)
\label{projectors}
\eeq
project the spin along a polarization vector $n^\mu$ 
satisfying $n^2 = -1$ and $n\cdot p = 0$
but otherwise having arbitrary orientation.
The freedom in the choice of 
the unit spacelike transverse vector $n^\mu$ 
reflects the spin degeneracy of the eigenspinors.
However,
in the presence of perturbative Lorentz violation,
the breaking of spin degeneracy for $\BD\neq 0$
implies that each solution to the dispersion relation
becomes an eigenmode having a definite spin polarization.
The polarization projection operators $P_\pm$ 
must therefore involve a vector $n^\mu$ 
with a definite orientation. 
Next,
we obtain an approximation representation of $n^\mu$.

The birefringent term involving $\pm \BD$ can be isolated
from the modified Dirac equation
at leading order in Lorentz violation
by acting on the left with the operator 
$(p+\Vhat)\cdot\ga + (\m-\Shat)$.
Using both the modified Dirac equation 
and the result \rf{drapprox}
in the form $[(p+\Vhat)^2 - (\m-\Shat)^2]^2 \approx 4\BD^2$
permits the elimination of terms at second order 
in Lorentz-violating operators.
This generates the equation 
\beq
[\pm \BD -\ga_5(p\cdot\Ahat -\m \Ahat\cdot\ga
-p\cdot\Tdual\cdot\ga)]\ps \approx 0 ,
\eeq
which for $\BD\neq 0$ motivates the definition
\beq
P_\pm \approx \half\Bigg(1 \pm \fr{\ga_5\big(
p\cdot\Ahat -\m\Ahat\cdot\ga
-p\cdot\Tdual\cdot\ga\big)}{\BD}\Bigg) .
\label{projdef}
\eeq
A short calculation reveals that
\beq
[\ga_5 ( p\cdot\Ahat -\m \Ahat\cdot\ga
-p\cdot\Tdual\cdot\ga)]^2 = \BD^2 ,
\eeq
ensuring that $P_\pm$ are 
indeed orthogonal projection operators.

To express the projectors \rf{projdef}
in the form \rf{projectors},
we use $\m\ps\approx p\cdot\ga\ps$
and thereby identify 
the spacelike vector $n^\mu$ as
\beq
n^\mu \approx
\fr{p\cdot\Ahat p^\mu - \m^2\Ahat^\mu
+\m\Tdual\pt{}^{\mu\nu} p_\nu}{\m\BD}
\equiv
\fr 1 {\BD} \nv^\mu,
\label{polvec}
\eeq
which satisfies $n^2 = -1$ and $n\cdot p \approx 0$,
as required.
The spacelike vector 
\beq
\nv^\mu = ( \Tdual^{\mn} + \tfrac 1 \m p^{[\mu}\Ahat^{\nu]} ) p_\nu
= \Tdual_\eff^{\mn} p_\nu
\label{N}
\eeq
obeying $\nv^2 = -\BD^2$ and $\nv \cdot p \approx 0$
is introduced for notational convenience in what follows.
The expression for $n^\mu$,
which is at zeroth order in Lorentz violation,
fixes the dominant polarization required
for a solution to approximate the exact eigenspinor. 
Note that if $\BD=0$ the derivation breaks down,
but the spin degeneracy is then restored
and so $n^\mu$ can be approximated 
as a unit spacelike transverse vector,
as usual.
Note also that the subscripts on the projections $P_\pm$
correspond to the signs in the modified dispersion relation \rf{drapprox2},
so polarizing a fermion along $n^\mu$ increases the energy 
while the opposite polarization decreases it.

\section{Hamiltonian}
\label{Hamiltonian}

The construction of the exact hamiltonian 
associated with the full theory \rf{lag}
is complicated by the higher-order time derivatives that appear.
For most practical applications,
however,
it suffices to obtain an effective hamiltonian
that describes correctly the behavior at leading order in Lorentz violation.
We present here a perturbative derivation of the hamiltonian
via a generalization of the standard approach,
and we extract the relativistic combinations of coefficients 
that it contains.

\subsection{Construction}
\label{Construction}

The goal of the standard approach to constructing the hamiltonian 
is to find a unitary transformation $U= U(\pvec)$ 
converting the modified Dirac equation \rf{moddirac} to the form
\beq
U\ga_0(p\cdot\ga - \m + \Qhat)U^\dag U\ps 
= (\eom - H) U\ps = 0 ,
\label{udirac}
\eeq
where $\eom \equiv p_0$ 
and the $4\times 4$ relativistic hamiltonian $H$ 
is block diagonal or `even' 
with vanishing $2\times 2$ off-diagonal `odd' blocks.
This decouples the positive and negative energy states,
and the diagonal blocks give the $2\times 2$ relativistic hamiltonians 
describing particles and antiparticles.
We adopt the chiral representation,
in which the matrices $\ga^\mu$ are block off diagonal,
so we seek $U$ such that Eq.\ \rf{udirac}
involves only an even number of $\ga^\mu$ matrices.
Since the Lorentz violation is perturbative,
it is useful to write
\beq
H= H_0+\de H, 
\eeq
where $H_0 = \ga_0(\pvec\cdot\gavec+\m)$
is the usual $4\times 4$ Dirac hamiltonian for the Lorentz-invariant case 
and $\de H$ contains the Lorentz-violating modifications.

Consider first the usual Lorentz-invariant case
with $\Qhat=0$ and $\de H=0$.
An appropriate transformation $U=VW=WV$ is the product 
of the two commuting transformations
\beq
V = \fr{1+\ga_0\ga_5}{\sqrt{2}},
\quad 
W(\pvec) = \fr{\eomz + \m + \pvec\cdot\gavec}{\sqrt{2\eomz(\eomz+\m)}} ,
\label{usualu}
\eeq
with $\eomz = \sqrt{\pvec^2 + \m^2}>0$.
Direct calculation shows that
this transformation gives the expected block-diagonal hamiltonian
\beq
H_0 = -\ga_5\eomz
= \begin{pmatrix}
  \eomz & 0 \\
  0 & -\eomz
\end{pmatrix} .
\eeq
The upper $2\times 2$ block describes positive-energy particles 
with hamiltonian $h_0 = \eomz$,
while after the usual reinterpretation
the lower negative-energy block gives 
the hamiltonian $\ol h_0 = \eomz$ for positive-energy antiparticles.

In the Lorentz-violating case,
if $\Qhat$ is nonzero but contains only
the operators $\Shat$ and $\Vhat^\mu$,
then the same procedure can be used to perform the block diagonalization.
It suffices to replace 
$p_\mu$ with $p_\mu+\Vhat_\mu$ and $\m$ with $\m-\Shat$ 
in the transformation \rf{usualu}.
In contrast,
the general case involving also nonzero
$\Phat$, $\Ahat^\mu$, and $\That^\mn$ is challenging.
However,
a perturbative treatment can be adopted 
to implement the block diagonalization 
at leading order in Lorentz violation.

To zeroth order,
the transformation $U$ is given by the product $VW$.
So, we start by applying this,
\beq
VW\ga_0(p\cdot\ga - \m + \Qhat) W^\dag V^\dag 
= \eom  + \ga_5\eomz  + VW\ga_0\Qhat W^\dag V^\dag \ .
\eeq
The Lorentz-invariant terms are block diagonal,
but the last term contains both even and odd parts.
The even part of any matrix $M$ can be extracted
by applying the even matrix $\ga_5$ to give
$M_{\rm even} = (M + \ga_5 M \ga_5)/2$. 
To remove the odd part at first order in Lorentz violation
we can therefore modify $U$ by an additional small transformation,
\beq
U = \Big(1+\fr{1}{4\eomz}\big[\ga_5,VW\ga_0\Qhat W^\dag V^\dag\big]\Big)VW .
\label{Utrans}
\eeq
This gives 
\beq
U\ga_0(p\cdot\ga - \m + \Qhat) U^\dag 
= \eom  + \ga_5\eomz  + \big(VW\ga_0\Qhat W^\dag V^\dag\big)_{\rm even}
\eeq
at first order in Lorentz violation.
We can now identify the leading-order block-diagonal
Lorentz-violating hamiltonian as
\beq
\de H = - \big(VW \ga_0\Qhat W^\dag V^\dag\big)_{\rm even}.
\eeq
Substituting for $\Qhat$ using Eq.\ \rf{qhatsplit}
and performing some explicit calculations,
we obtain the result
\bea
\de H &=& 
\fr{1}{\eomz}\big[
\m\Shat \ga_5
-\eomz \Vhat^0 -\Vhat^j p^j \ga_5
+ \Ahat^0 p^j\ga^j\ga_0
\nn\\
&& 
+ \m \Ahat^j\ga^j\ga_0\ga_5 
+ \Ahat^jp^j p^k\ga^k \ga_0\ga_5/ (\eomz+\m)
\nn\\
&& 
+i p^j \That^{0k} \ga ^j \ga^k 
+i \That^{0j}p^j 
- \eomz \Tdual^{0j} \ga^j \ga_0
\nn\\
&& 
+ \Tdual^{0j} p^j  p^k \ga^k \ga_0 / (\eomz+\m)
\big] .
\label{dH}
\eea
The upper $2\times 2$ block of this operator represents 
the leading-order perturbation $\de h$ 
to the positive-energy hamiltonian for particles,
\beq
\de h = \fr{ \De + \svvec\cdot\sivec}{\eomz},
\label{h}
\eeq
where
\beq 
\De  = -\m\Shat - \eomz \Vhat^0 + p^j \Vhat^j
= - p_\mu \Vhat_\eff^\mu
\label{Deltaterm}
\eeq
and the spin dependence is controlled by
\bea
\sv^j &=& 
\Ahat^0 p^j - \m\Ahat^j - \Ahat^kp^k p^j/(\eomz+\m)
\nn\\
&& 
- \eomz\Tdual^{0j} - \Tdual^{jk} p^k + \Tdual^{0k}p^k p^j/(\eomz+\m) 
\nn\\
&=& - \eomz\Tdual^{0j}_\eff 
- \Tdual^{jk}_\eff p^k + \Tdual^{0k}_\eff p^k p^j/(\eomz+\m) .
\quad
\label{Sigma}
\eea
These results reduce to those established in Ref.\ \cite{kla}
for operators of minimal dimension $d=3$ and $d=4$,
as expected.
We emphasize that the $2\times 2$ hamiltonian 
\beq
h=h_0+\de h
\eeq
is fully relativistic.

After reinterpretation,
the lower $2\times 2$ block of \rf{dH} gives the change $\de\ol h$
to the positive-energy hamiltonian for antiparticles:
\beq
\de \ol h = \fr{ \ol \De + \svvec\cdot\ol\sivec}{\eomz},
\label{hbar} 
\eeq
where
\beq
\ol\De = -\m\Shat + \eomz \Vhat^0 + p^j \Vhat^j
\eeq
and 
\bea
\ol\sv^j &=&
\Ahat^0 p^j + \m\Ahat^j + \Ahat^kp^k p^j/(\eomz+\m)
\nn\\
&&
- \eomz\Tdual^{0j} + \Tdual^{jk} p^k + \Tdual^{0k}p^k p^j/(\eomz+\m) .
\qquad
\label{Sigmabar}
\eea
Note that in the Lorentz-violating terms we can take
$p_0 \approx \eomz$ for particles
and $p_0 \approx -\eomz$ for antiparticles
because corrections to these approximations 
contribute only at second order.
Since the physical antiparticle 3-momentum is $-\pvec$, 
the corresponding physical 4-momentum can be taken to be $-p_\mu$.
This implies that the antiparticle hamiltonian 
\beq
\ol h = \ol h_0 + \de \ol h
\eeq
can be obtained from $h$ 
by changing the sign of all coefficients for CPT-odd operators,
as expected.

The Lorentz-violating portion of the transformation \rf{Utrans}
can be expressed in an alternative form
by commuting $VW$ through to the right.
This gives 
\beq
U = VW\Big(1 -\fr{1}{4\eomz^2}\big[H_0 , \ga_0\Qhat\, \big]\Big) .
\label{Utrans2}
\eeq
We then find
\bea
U\ga_0(p\cdot\ga - \m + \Qhat) U^\dag 
\nn\\
&&
\hskip -90pt
= VW\big(\eom - H_0
+ \La_+ \ga_0\Qhat \La_+ 
+ \La_- \ga_0\Qhat \La_- 
\big) W^\dag V^\dag,
\nn\\
\eea
where $\La_\pm = (1\pm H_0/\eomz)/2$
are the usual projection operators for energy.
This equation reveals that the net effect 
of the Lorentz-violating part of $U$ 
is to remove the portions of $\Qhat$ mixing 
the usual particle and antiparticle states.

\subsection{Coefficients}
\label{Coefficients}

The explicit nature of the terms \rf{Deltaterm} and \rf{Sigma}
in the perturbation hamiltonian $\de h$ 
obscures the relativistic combinations of coefficients
from which they are formed.
A more elegant form for $\de h$ that displays these combinations
can be obtained using the relativistic polarization vector $\nv^\mu$
defined in Eq.\ \rf{polvec}.

The spin vector $\sv^j$ is related to $\nv^\mu$ by
\bea
\sv^j &=& \nv^j - \fr{\nv^0 p^j} {\eomz + \m}
= \nv^j - \fr {\nv^kp^kp^j}{\eomz (\eomz + \m)}
\nn\\
&=& \nv_\perp^j + \fr{\m}{\eomz} \nv_\parallel^j ,
\label{svec}
\eea
where $\nvec_\perp$ and $\nvec_\parallel$ 
are the components of $\nvec$ perpendicular and parallel to $\pvec$, 
respectively.
Note that both $\nvec_\perp$ and $\m\nvec_\parallel$
remain finite even in the massless limit.
The magnitude of the spin vector is $|\svvec| = \BD$,
so the spin-dependent energy shifts are 
$\BD/\eomz$ for spin along $\svvec$
and $-\BD/\eomz$ for spin opposite $\svvec$,
as expected.

To gain further insight,
consider a massive particle in its rest frame
with $\nv^0=0$ and $\nvec^2 = \BD^2$,
and introduce the rest-frame polarization unit vector $\nvec'$.
Boosting to an arbitrary frame then gives
\beq
\nv^0 = \fr{\pmag}{\m} |\nvec'_\parallel| , \quad
\nvec = \nvec'_\perp + \fr{\eomz}{\m} \nvec'_\parallel ,
\eeq
where $\nvec'_\parallel$ and $\nvec'_\perp$ 
are the projections of $\nvec'$ parallel and perpendicular to $\pvec$,
respectively.
Comparing to Eq.\ \rf{svec} reveals that 
$\svvec$ is the rest-frame $\nvec$ vector,
\beq
\svvec = \nvec' .
\eeq
Note that $\nvec'$ depends on $p_\mu$ because the required boost 
varies with $p_\mu$.

The above considerations permit us to write 
$\svvec\cdot\sivec$ as  
\beq
\svvec\cdot\sivec = -\nv^\mu \ta_\mu ,
\label{sisi}
\eeq
where
\beq
\ta^0 = \fr{\pvec\cdot\sivec}{\m} , 
\quad
\ta^j = \si^j + \fr{\pvec \ta^0}{\eomz + \m} .
\eeq
The perturbation hamiltonian \rf{h} therefore takes the form
\beq
\de h = \fr{\De -\nv^\mu\ta_\mu}{\eomz},
\label{hn}
\eeq
showing that the spin-dependent Lorentz violation
is fixed by the relativistic polarization vector $\nv^\mu$ 
given in Eq.\ \rf{polvec}.

We can now expand $\De$ and $\nv^\mu$ in powers of momentum $p^\mu$
to extract the effective coefficients for Lorentz violation
that appear in the hamiltonian.
In practice,
it is convenient to split $\De$ and $\nv^\mu$ 
into CPT odd and CPT even pieces
for this purpose.

Expanding $\De$ yields 
\beq
\De = \sum_d \De^{(d)\al_1 \ldots\al_{d-2}} p_{\al_1} \ldots p_{\al_{d-2}},
\label{Delta}
\eeq
where even and odd $d$ are associated with
CPT-even and CPT-odd Lorentz violations,
respectively,
and the coefficients $\De^{(d)\al_1 \ldots\al_{d-2}}$
have mass dimension $4-d$.
Separating the CPT-even and CPT-odd parts 
and substituting the definitions \rf{effcompops}
into the expression \rf{Delta} gives
\bea
\De_{\rm odd} &\equiv& \ahat_\eff^\mu p_\mu
= \sum_d a_\eff^{(d)\al_1\ldots\al_{d-2}} 
p_{\al_1} \ldots p_{\al_{d-2}} ,
\nn\\
\De_{\rm even} &\equiv& - \chat_\eff^\mu p_\mu
= -\sum_d c_\eff^{(d)\al_1\ldots\al_{d-2}} 
p_{\al_1} \ldots p_{\al_{d-2}} ,
\qquad
\label{aceff}
\eea
where the effective coefficients are given
in terms of fundamental coefficients by Eq.\ \rf{effcoeffcomps}.

Similarly, 
expanding $\nv^\mu$ yields
\beq
\nv^\mu = 
\sum_d \nv^{(d)\mu\al_1 \ldots\al_{d-2}} p_{\al_1} \ldots p_{\al_{d-2}} ,
\label{capnmu}
\eeq
where now even and odd $d$ are associated with 
CPT-odd and CPT-even Lorentz violations,
respectively,
with the coefficients $\nv^{(d)\mu\al_1 \ldots\al_{d-2}}$
having mass dimension $4-d$.
Note that constant $\nv^\mu$ is forbidden 
by the restriction $p\cdot \nv=0$.
Separating the CPT-even and CPT-odd parts
and combining the definitions \rf{effcompops}
and the result \rf{N} gives
\bea
\nv^\mu_{\rm odd} 
&\equiv& \gdual^\mn_\eff p_\nu
= \sum_d \gt_\eff^{(d)\mu\al_1 \ldots\al_{d-2}} 
p_{\al_1} \ldots p_{\al_{d-2}} ,
\nn\\
\nv^\mu_{\rm even} 
&\equiv& -\Hdual^\mn_\eff p_\nu
= -\sum_d \Ht_\eff^{(d)\mu\al_1 \ldots\al_{d-2}} 
p_{\al_1} \ldots p_{\al_{d-2}} ,
\nn\\
\label{gheff}
\eea
where again the effective coefficients are given
in terms of fundamental coefficients by Eq.\ \rf{effcoeffcomps}.

The above analysis reveals that 
the perturbative hamiltonian $\de h$ in Eq.\ \rf{h}
can conveniently be split into four pieces
according to 
\bea
\de h &=& h_a + h_c + h_g + h_H ,
\nn\\
&=& \fr 1{\eomz}
(\ahat_\eff^\nu - \chat_\eff^\nu 
- \gdual^\mn_\eff \ta_\mu + \Hdual^\mn_\eff \ta_\mu ) p_\nu,
\label{pham}
\eea
where the explicit expansions for 
$\ahat_\eff^\mu$,
$\chat_\eff^\mu$, 
$\gdual^\mn_\eff$,
and $\Hdual^\mn_\eff$
are given by Eqs.\ \rf{aceff} and \rf{gheff},
respectively.
Each of the four component hamiltonians
is uniquely specified by spin and CPT properties:
the spin-independent terms $h_a$ and $h_c$
are CPT-odd and CPT-even,
respectively,
as are the spin-dependent terms $h_g$ and $h_H$.
Note that the structure of the results obtained above 
is compatible with the discussion in Sec.\ \ref{Field redefinitions}
concerning field redefinitions and physical observables.

\section{Spherical decomposition}
\label{Spherical decomposition}

The complexity of the two-component perturbative hamiltonian \rf{pham}
and the appearance of coefficients with numerous indices 
make a general analysis of physical implications 
unwieldy for arbitrary $d$.
Some of the difficulties can be alleviated by
performing a spherical-harmonic decomposition of the hamiltonian.
For example,
a typical experimental application 
involves a transformation from a noninertial laboratory frame 
to the canonical Sun-centered inertial frame
\cite{km02,sunframe,tables},
which is generically dominated by rotations
and is therefore simpler in spherical basis.
For each $d$,
the spherical-harmonic decomposition
yields a set of coefficients
equivalent to those introduced in Sec.\ \ref{Basics}
but having comparatively simple rotation properties.
This permits a systematic classification
of the coefficients affecting the dynamics
and is also advantageous because rotation violations 
are a key signature of Lorentz violation.

\subsection{Basics}
\label{sphericalBasics}

Since the hamiltonian \rf{pham} is expressed in momentum space,
the relevant spherical coordinates also lie in this space.
We can introduce spherical polar angles $\th$, $\ph$ 
via the unit 3-momentum vector $\phat = \pvec/\pmag$
written in the form $\phat = (\sin\th\cos\ph,\sin\th\sin\ph,\cos\th)$.
Rotation scalars can then be expanded 
in terms of the usual spherical harmonics
$\syjm{0}{jm}(\phat)\equiv Y_{jm}(\th,\ph)$.
However,
the expansion of rotation tensors 
requires some form of generalized spherical harmonics.
We adopt here the spin-weighted spherical harmonics 
$\syjm{s}{jm}(\phat)\equiv\syjm{s}{jm}(\th,\ph)$,
which permit the spherical decomposition of tensors in the helicity basis.
The spin weight $s$ of an irreducible tensor 
is defined as the negative of its helicity
and is limited by $|s| \leq j$.
A summary of properties of the spin-weighted spherical harmonics
is given in Appendix A of Ref.\ \cite{km09}.

In the perturbative hamiltonian \rf{pham},
$h_a$ and $h_c$ transform as scalars under rotations,
while $h_g$ and $h_H$ are spin dependent
through the quantity $\svvec\cdot\sivec = -\nv^\mu \ta_\mu$ 
and so have nontrivial rotation properties.
To perform the expansion in spin-weighted spherical harmonics,
we therefore require the decomposition of $\svvec\cdot\sivec$
in the helicity basis.
The helicity basis vectors are defined as
$\ephat_r = \ephat^r = \phat$
and $\ephat_\pm = \ephat^\mp = (\thhat \pm i\phhat)/\sqrt{2}$,
where $\thhat$ and $\phhat$ are the usual unit vectors
associated with the polar angle $\th$ and azimuthal angle $\ph$.
The helicity decomposition is
\beq
\svvec\cdot\sivec 
= \sv_w \si^w = \sv^w \si_w = \sv^- \si_- + \sv^r \si_r + \sv^+ \si_+ ,
\eeq
where the repeated index $w$ is summed over $w=+,r,-$,
and $\si_w = \ephat_w\cdot\sivec$, $\si^w = \ephat^w\cdot\sivec$.
The component $\sv_r = \ephat_r\cdot\svvec$ is a rotational scalar
with spin weight zero and can therefore be expanded 
in the usual spherical harmonics $\syjm{0}{jm}(\phat)$.
The components $\sv_\pm = \ephat_\pm\cdot\svvec$
have spin weight $s=\pm1$
and can be expanded in the harmonics $\syjm{\pm1}{jm}(\phat)$,
while the components $\sv^\pm = \ephat^\pm\cdot\svvec = \sv_\mp$
have helicity $\pm1$.
The Pauli matrices in the helicity basis are
\bea
\si_r = \si^r &=& \begin{pmatrix}
\cos\th          & \sin\th e^{-i\ph}\\
\sin\th e^{i\ph} & -\cos\th
\end{pmatrix} ,
\nn\\
\si_\pm = \si^\mp 
&=& \fr{1}{\sqrt{2}}\begin{pmatrix}
-\sin\th          & (\cos\th\pm 1) e^{-i\ph}\\
(\cos\th\mp 1) e^{i\ph} & \sin\th \end{pmatrix} .
\qquad 
\eea

Up to constants,
$\si_r$ is the helicity operator
and $\si_\pm$ are helicity ladder operators.
To see this,
consider the special frame in which 
$\thhat = \xhat, \phhat=\yhat, \phat = \zhat$
and so 
\beq
\si_r = \begin{pmatrix}1&0\\0&-1\end{pmatrix},
\quad
\si_+ = \sqrt2 \begin{pmatrix}0&1\\0&0\end{pmatrix},
\quad
\si_- = \sqrt2 \begin{pmatrix}0&0\\1&0\end{pmatrix}.
\eeq
This shows that $\si_+$ raises the helicity and $\si_-$ lowers it.
Acting with $\svvec\cdot\sivec$ on a spinor $\ph$ 
with components $(\ph_\uparrow, \ph_\downarrow)$ 
having helicities $(+1/2,-1/2)$, respectively,
gives
\beq
\svvec\cdot\sivec 
\begin{pmatrix}\ph_\uparrow\\ \ph_\downarrow\end{pmatrix}
=\begin{pmatrix}
\sv^r \ph_\uparrow + \sqrt2 \sv^+ \ph_\downarrow\\
-\sv^r\ph_\downarrow + \sqrt2 \sv^- \ph_\downarrow
\end{pmatrix} .
\eeq
The first component maintains its positive helicity
because $\sv^r \ph_\uparrow$ is the product of
objects with helicity $0$ and $+1/2$,
while $\sv^+ \ph_\downarrow$ is the product of
objects with helicity $+1$ and $-1/2$.
Similarly,
the second component remains an object of helicity $-1/2$.
We can conclude that $\sv^r = \sv_r$ 
generates helicity-dependent effects without changing the helicity,
$\sv^+=\sv_-$ is associated with a raising of helicity,
and $\sv^-=\sv_+$ is associated with a lowering of helicity.

\subsection{Decomposition}
\label{Decomposition}

We can now proceed with the spherical decomposition
of the perturbative hamiltonian \rf{pham}.
The components $h_a$ and $h_c$ 
are rotational scalars and so can be expanded 
in the usual spherical polar coordinates as
\bea
h_a &=& 
\sum_{dnjm} \eomz^{d-3-n} \pmag^n\,
\syjm{0}{jm}(\phat)\, \acoef{d}{njm} ,
\nn\\
h_c &=& 
- \sum_{dnjm} \eomz^{d-3-n} \pmag^n\,
\syjm{0}{jm}(\phat)\, \ccoef{d}{njm} .
\label{hc}
\eea
In contrast,
the spin-dependent component hamiltonians $h_g$ and $h_H$
transform nontrivially under rotations 
and must therefore first be separated 
into spin-weighted components,
\beq
h_g = (h_g)_w \si^w ,
\quad
h_H = (h_H)_w \si^w .
\label{swc}
\eeq
At the end of this subsection, 
we show that these components have expansions 
\bea
(h_g)_r 
&=& - \m \sum_{dnjm} \eomz^{d-4-n}\pmag^n \, \syjm{0}{jm}(\phat) 
\nn\\
&&
\hskip 110pt
\times 
(n+1)\gzBcoef{d}{njm} ,
\nn\\
(h_g)_\pm
&=& \sum_{dnjm} \eomz^{d-3-n}\pmag^n\, \, \syjm{\pm 1}{jm}(\phat)
\nn\\
&&
\times 
\Big[\pm\sqrt{\frac{j(j+1)}{2}} \gzBcoef{d}{njm}
\pm\goBcoef{d}{njm} + i \goEcoef{d}{njm} \Big] ,
\nn\\
\nn\\
(h_H)_r 
&=& \m \sum_{dnjm} \eomz^{d-4-n}\pmag^n\, \syjm{0}{jm}(\phat) 
\nn\\
&&
\hskip 110pt
\times 
(n+1)\HzBcoef{d}{njm} ,
\nn\\
(h_H)_\pm
&=& - \sum_{dnjm} \eomz^{d-3-n}\pmag^n\, \, \syjm{\pm 1}{jm}(\phat) 
\nn\\
&&
\hskip-10pt
\times 
\Big[\pm\sqrt{\frac{j(j+1)}{2}} \HzBcoef{d}{njm}
\pm\HoBcoef{d}{njm} + i \HoEcoef{d}{njm} \Big] .
\nn\\
\label{SiH}
\eea
The full perturbative hamiltonian \rf{pham}
is therefore given by the expansion
\bea
\de h &=& 
h_a + h_c 
+ (h_g)_+ \si^+ + (h_g)_r \si^r + (h_g)_- \si^- 
\nn\\
&&
+ (h_H)_+ \si^+ + (h_H)_r \si^r + (h_H)_- \si^- ,
\qquad
\label{helh}
\eea
where the component hamiltonians
are given by Eqs.\ \rf{hc} and \rf{SiH}. 

\begin{table*}[t]
\renewcommand{\arraystretch}{1.5}
\begin{tabular*}{\textwidth}{*{7}{@{\extracolsep{\fill}}c}}
Coefficient        & CPT  & Parity type & $d$            & $n$              & $j$                       & Number \\\hline\hline
$\acoef{d}{njm}$   & odd  & $E$    & odd, $\geq3$   & $0,1,\ldots,d-2$ & $n,n-2,n-4,\ldots\geq0$   & $\tfrac16 (d+1)d(d-1)$ \\
$\ccoef{d}{njm}$   & even & $E$    & even, $\geq 4$ & $0,1,\ldots,d-2$ & $n,n-2,n-4,\ldots\geq0$   & $\tfrac16 (d+1)d(d-1)$ \\
$\gzBcoef{d}{njm}$ & odd  & $B$    & even, $\geq 4$ & $0,1,\ldots,d-3$ & $n+1,n-1,n-3,\ldots\geq0$ & $\tfrac16 (d+1)d(d-1)-1$ \\
$\goBcoef{d}{njm}$ & odd  & $B$    & even, $\geq 4$ & $2,3,\ldots,d-2$ & $n-1,n-3,n-5,\ldots\geq1$ & $\tfrac16 (d-2)(d^2-d-3)$ \\
$\goEcoef{d}{njm}$ & odd  & $E$    & even, $\geq 4$ & $1,2,\ldots,d-2$ & $n,n-2,n-4,\ldots\geq1$ & $\tfrac16 (d+2)d(d-2)$ \\
$\HzBcoef{d}{njm}$ & even & $B$    & odd, $\geq3$   & $0,1,\ldots,d-3$ & $n+1,n-1,n-3,\ldots\geq0$ & $\tfrac16 (d+1)d(d-1)-1$ \\
$\HoBcoef{d}{njm}$ & even & $B$    & odd, $\geq5$   & $2,3,\ldots,d-2$ & $n-1,n-3,n-5,\ldots\geq1$ & $\tfrac16 (d+1)(d-1)(d-3)$ \\
$\HoEcoef{d}{njm}$ & even & $E$    & odd, $\geq3$   & $1,2,\ldots,d-2$ & $n,n-2,n-4,\ldots\geq1$ & $\tfrac16 (d-1)(d^2+d-3)$ \\
\hline
$\afc{d}{n}$ & odd        & even   & odd, $\geq3$   & $0,2,4,\ldots,d-3$ & 0 & $\half(d-1)$ \\
$\cfc{d}{n}$ & even       & even   & even, $\geq4$  & $0,2,4,\ldots,d-2$ & 0 & $\half d$ \\
$\gfc{d}{n}$ & odd        & odd    & even, $\geq4$  & $1,3,5,\ldots,d-3$ & 0 & $\half (d-2)$ \\
$\Hfc{d}{n}$ & even       & odd    & odd, $\geq5$   & $1,3,5,\ldots,d-4$ & 0 & $\half(d-3)$\\
\hline
\end{tabular*}
\caption{
\label{spherical_coefs} 
Spherical coefficients for Lorentz violation. 
}
\end{table*}

The properties and index ranges of the eight sets of spherical coefficients 
appearing in these expansions
are summarized in Table \ref{spherical_coefs}.
All coefficients have mass dimension $4-d$.
The first column of the table
lists the coefficients.
The second column specifies the CPT handedness of the corresponding operators.
The third column gives the behavior of the operators under parity,
where operators with $E$-type parity acquire a sign $(-1)^j$ 
and those with $B$-type parity acquire a sign $(-1)^{j+1}$. 
The next three columns lists the allowed ranges of $d$, $n$, and $j$,
while the final column provides the number of independent coefficients
appearing for each $d$.

The coefficients given in the expansions \rf{hc} and \rf{SiH}
and listed in Table \ref{spherical_coefs}
comprise the set of observable quantities 
at leading order in Lorentz violation.
Each set of coefficients $\K_{jm}$ obeys
the complex conjugation relation
\beq
\K_{jm}^* = (-1)^m \K_{j(-m)} ,
\eeq
which stems from the reality of the underlying tensors in momentum space
and ultimately from the hermiticity 
of the hamiltonian \rf{pham} and the theory \rf{lag}.
The coefficients have comparatively simple properties under rotations,
which can be implemented using the standard Wigner rotation matrices 
in parallel with the treatments for the photon and neutrino sectors
given in Sec.\ V of Ref.\ \cite{km09}
and Sec.\ VI of Ref.\ \cite{km12}.
As an example,
the relation between coefficients $\K_{jm}^{\rm lab}$ 
in a standard laboratory frame
with $x$ axis pointing south and $y$ axis pointing east
to coefficients $\K_{jm}$ in the canonical Sun-centered frame
\cite{km02,sunframe,tables}
is 
\beq
\K_{jm}^{\rm lab}
= \sum_{m'} e^{im'\om_\oplus T_\oplus} 
d^{(j)}_{mm'}(-\ch) \K_{jm'} ,
\label{coefrot}
\eeq
where 
$\om_\oplus$ is the sidereal rotation frequency of the Earth,
$T_\oplus$ is the sidereal time,
the quantities $d^{(j)}_{mm'}$ are the `little' Wigner matrices
given in Eq.\ (136) of Ref.\ \cite{km09},
and $\ch$ is the colatitude of the laboratory in the northern hemisphere.
This expression only involves a linear combination mixing
the azimuthal components labeled by $m'$.

The subset of isotropic coefficients can be identified by imposing $j=m=0$
in the spherical-harmonic expansion of the hamiltonian \rf{pham}.
In this limit,
the helicity-flipping pieces of the hamiltonian vanish.
This is because helicity $\pm 1$ is incompatible with $j=0$ 
or, equivalently, 
because the spin-weight is limited by $|s|\leq j$.
As a result,
the perturbative isotropic hamiltonian takes the form 
\beq
\de \ring{h} = \ring{h}_a + \ring{h}_c 
+ (\ring{h}_g)_r \si^r + (\ring{h}_H)_r \si^r ,
\label{phami}
\eeq
where
\bea
\ring{h}_a &=& 
\sum_{dn} \eomz^{d-3-n} \pmag^n \afc{d}{n} ,
\nn\\
\ring{h}_c &=& 
-\sum_{dn} \eomz^{d-3-n} \pmag^n \cfc{d}{n} ,
\nn\\
(\ring{h}_g)_r &=& 
-\m \sum_{dn} \eomz^{d-4-n} \pmag^n \gfc{d}{n} ,
\nn\\
(\ring{h}_H)_r &=& 
\m \sum_{dn} \eomz^{d-4-n} \pmag^n \Hfc{d}{n} .
\label{SiH_fc}
\eea
In these expressions,
the isotropic coefficients 
are related to the spherical coefficients through
\bea
\afc{d}{n} &=& \fr{1}{\sqrt{4\pi}} \acoef{d}{n00} ,
\nn\\
\cfc{d}{n} &=& \fr{1}{\sqrt{4\pi}} \ccoef{d}{n00} ,
\nn\\
\gfc{d}{n} &=& \fr{1}{\sqrt{4\pi}} (n+1) \gzBcoef{d}{n00} ,
\nn\\
\Hfc{d}{n} &=& \fr{1}{\sqrt{4\pi}} (n+1) \HzBcoef{d}{n00} .
\eea
These relations give equivalent representations 
for the spherical coefficients with $j=0$ 
listed in Table \ref{spherical_coefs}.
All the isotropic coefficients have mass dimension $4-d$.
Their index ranges and counting are summarized 
in Table \ref{spherical_coefs}.
Note that the result \rf{lorinv} implies exactly one linear combination
of $\cfc{d}{n}$ at each even $d$ controls a Lorentz-invariant operator.

To illustrate the connection
between the cartesian and isotropic coefficients 
in the context of the Lagrange density \rf{lag},
we can consider the explicit form of the effective isotropic theory
for the first few dimensions $d=3,4,5,6$.
At $d=3$ only one term exists, 
\beq
\Qhat_\eff^{(3)} = -a_\eff^{(3)0} \ga_0 \equiv -\afc{3}{0} \ga_0,
\label{three}
\eeq
representing an isotropic CPT-violating operator.
Note, however,
that the phase redefinition \rf{phaseredef} can be used to show 
this term has no observable effects,
as discussed in Sec.\ \ref{Field redefinitions}.
At $d=4$ there are three independent isotropic terms,
given by 
\bea
\Qhat_\eff^{(4)} &=& 
c_\eff^{(4)00} p^0\ga^0 + \tfrac13 c_\eff^{(4)jj} p^k \ga^k
+ \tfrac13 i \gt_\eff^{(4)0jj} p^k \ga_5\si^{0k}
\nn\\
&\equiv& \cfc{4}{0} p^0\ga^0 + \cfc{4}{2} p^k \ga^k
- i \gfc{4}{1} p^k \ga_5\si^{0k} .
\label{four}
\eea
Since $d$ is even,
one combination of the coefficients $\cfc{4}{n}$ must be associated
with a Lorentz-invariant operator,
and it is 
\beq
c_\eff^{(4)00}-c_\eff^{(4)jj}
=\cfc{4}{0}-3\cfc{4}{2}.
\eeq
At $d=5$, 
the theory also contains three independent isotropic terms,
\bea
\Qhat_\eff^{(5)} &=& 
-a_\eff^{(5)000} p^0p^0\ga^0 
- \tfrac13 a_\eff^{(5)0jj} \big(p^kp^k \ga^0 + 2p^0p^k \ga^k\big)
\nn\\
&&
- \tfrac23 i \Ht_\eff^{(5)0j0j} p^0 p^k \ga_5\si^{0k}
\nn\\
&\equiv&
- \afc{5}{0} p^0p^0\ga^0 
- \tfrac13 \afc{5}{2} \big(p^kp^k \ga^0 + 2p^0p^k \ga^k\big)
\nn\\
&&
+ i \Hfc{5}{1} p^0 p^k \ga_5\si^{0k} ,
\label{five}
\eea
all of which are Lorentz violating.
This is the lowest dimension $d$ at which 
the effective coefficients $\Hfc{d}{n}$ appear. 
Finally,
at $d=6$ there are five isotropic terms,
\bea
\Qhat_\eff^{(6)} &=& 
c_\eff^{(6)0000} p^0p^0p^0\ga^0 
\nn\\
&&
+ c_\eff^{(6)00jj} \big(p^0p^kp^k\ga^0+ p^0p^0p^k\ga^k\big)
\nn\\
&&
+ \tfrac15 c_\eff^{(6)jjkk} p^lp^lp^n\ga^n
\nn\\
&&
+i \gt_\eff^{(6)0j00j} p^0p^0p^k \ga_5\si^{0k}
\nn\\
&&
+\tfrac{1}{5}  i \gt_\eff^{(6)0jjkk} p^lp^lp^n \ga_5\si^{0n}
\nn\\
&\equiv& 
\cfc{6}{0} p^0p^0p^0\ga^0 
+ \tfrac12 \cfc{6}{2} \big(p^0p^kp^k\ga^0+ p^0p^0p^k\ga^k\big)
\nn\\
&&
+ \cfc{6}{4} p^lp^lp^n\ga^n
- i \gfc{6}{1}  p^0p^0p^k \ga_5\si^{0k}
\nn\\
&&
- i \gfc{6}{3} p^lp^lp^n \ga_5\si^{0n}.
\label{six}
\eea
At this dimension
another Lorentz-invariant trace appears,
associated with the coefficient combination
\beq
c_\eff^{(6)0000}-2c_\eff^{(6)00jj}+c_\eff^{(6)jjkk}
= \cfc{6}{0} - \cfc{6}{2} + 5\cfc{6}{4}.
\eeq
More generally,
both even dimensions $d=2k$ and odd dimensions $d=2k+1$
have $2k-1$ independent isotropic terms.
For even dimensions one combination is Lorentz invariant,
and the number of independent CPT-even and CPT-odd 
Lorentz-violating operators is the same.
For odd dimensions all terms are Lorentz violating,
and the CPT-odd Lorentz-violating operators
number one more than the CPT-even ones. 

The remainder of this section derives the results \rf{SiH}.
The reader uninterested in the derivation
can proceed directly to the discussion of dispersion and birefringence
in Sec.\ \ref{Dispersion and birefringence}.

The coefficients $\gt_\eff^{(d)\mu\nu\al_1\ldots\al_{d-3}}$
are antisymmetric in the first two indices
and symmetric in the remaining indices, 
and their appearance in the operator $\gdual_\eff^\mn p_\nu$ 
in conjunction with $p_\nu$
implies that they can be taken to vanish under antisymmetrization 
of any three indices.
The operator $\gdual_\eff^\mn$ 
therefore obeys the Maxwell-like equation
$\prt^{\la}_{\phantom{f}} \gdual_\eff^{\mn} 
+ \prt^{\mu}_{\phantom{f}} \gdual_\eff^{\nu\la}
+\prt^{\nu}_{\phantom{f}} \gdual_\eff^{\la\mu} = 0$,
which in turn constrains the coefficients in the spherical expansion.

To understand this constraint,
it is useful to define a pseudovector
$\E^j = \gdual_\eff^{j0}$
and a vector
$\B^j = -\ep^{jkl}\gdual_\eff^{kl}/2$,
in terms of which the constraint equation 
resembles the homogeneous Maxwell equations,
\beq
\del\times\cEvec + \prt_0 \cBvec = 0 , 
\quad \del\cdot\cBvec = 0 .
\label{max}
\eeq
Prior to imposing these constraint equations,
the spherical-harmonic expansion of the operators $\E^j$ and $\B^j$ 
can be written as
\bea
\E_r &=& \sum_{dnjm} \eomz^{d-3-n}\pmag^n\, Y_{jm}(\phat) 
\Edjm{0B}{d}{njm} ,
\nn\\
\E_\pm &=& \sum_{dnjm} \eomz^{d-3-n}\pmag^n\, \syjm{\pm 1}{jm}(\phat) 
\nn\\
&&
\hskip 30pt
\times
\big(\pm\Edjm{1B}{d}{njm} +i\Edjm{1E}{d}{njm}\big) ,
\nn\\
&&
\nn\\
\B_r &=& \sum_{dnjm} \eomz^{d-3-n}\pmag^n\, Y_{jm}(\phat) 
\Bdjm{0E}{d}{njm} ,
\nn\\
\B_\pm &=& \sum_{dnjm} \eomz^{d-3-n}\pmag^n\, \syjm{\pm 1}{jm}(\phat) 
\nn\\
&&
\hskip 30pt
\times
(\pm\Bdjm{1E}{d}{njm} +i\Bdjm{1B}{d}{njm}) .
\label{expa}
\eea
The equations \rf{max} imply interrelations 
between the six sets of coefficients in these expansions.

The first equation of Eqs.\ \rf{max}
yields two constraints on $E$-type coefficients
and one on $B$-type coefficients,
\bea
\Bdjm{0E}{d}{njm} &=& -\fr{\sqrt{2j(j+1)}}{n+2} \Bdjm{1E}{d}{njm} , 
\nn\\
\Edjm{1E}{d}{njm} &=& \fr{d-2-n}{n+1} \Bdjm{1E}{d}{(n-1)jm} , 
\nn\\
\Edjm{1B}{d}{njm} &=& -\fr{\sqrt{2j(j+1)}}{2(n+1)} \Edjm{0B}{d}{njm} 
\nn\\
&&
\hskip 40pt
- \fr{d-2-n}{n+1}\Bdjm{1B}{d}{(n-1)jm} .
\nn\\
\eea
The second of Eqs.\ \rf{max} ensures vanishing divergence of $\cBvec$ 
but provides no additional constraints.
Careful consideration of the index ranges of all the coefficients 
reveals that we can choose 
$\Edjm{0B}{d}{njm}$,
$\Bdjm{1B}{d}{njm}$, and
$\Bdjm{1E}{d}{njm}$ 
to be a set of independent coefficients.
Consequently, 
the expansions \rf{expa} become
\bea
\E_r &=& \sum_{dnjm} \eomz^{d-3-n}\pmag^n\, Y_{jm}(\phat) 
\Edjm{0B}{d}{njm} ,
\nn\\
\E_\pm &=& \sum_{dnjm} \eomz^{d-3-n}\pmag^n\, \syjm{\pm 1}{jm}(\phat) 
\nn\\
&&
\hskip -20pt
\times
\Big[ \mp\frac{\sqrt{2j(j+1)}}{2(n+1)} \Edjm{0B}{d}{njm} 
\nn\\
&&
+\frac{d-2-n}{n+1}\Big( \mp\Bdjm{1B}{d}{(n-1)jm}
+i \Bdjm{1E}{d}{(n-1)jm}\Big)
\Big] ,
\nn\\
\B_r &=& \sum_{dnjm} \eomz^{d-3-n}\pmag^n\, Y_{jm}(\phat) 
(-)\frac{\sqrt{2j(j+1)}}{n+2} \Bdjm{1E}{d}{njm} ,
\nn\\
\B_\pm &=& \sum_{dnjm} \eomz^{d-3-n}\pmag^n\,  \syjm{\pm 1}{jm}(\phat) 
\nn\\
&&
\hskip 20pt
\times
\Big(\pm\Bdjm{1E}{d}{njm} +i\Bdjm{1B}{d}{njm}\Big) .
\eea

The helicity-basis components of the hamiltonian $h_g$ 
can now be written as
\bea
(h_g)_r &=& \fr{\m}{\eomz^2} \punit^j \gdual_\eff^{j\nu} p_\nu 
= \fr{\m}{\eomz} \E_r , 
\nn\\
(h_g)_\pm &=& \fr{1}{\eomz} \epunit_\pm^j \gdual_\eff^{j\nu} p_\nu 
= \fr{1}{\eomz}\big(\eomz\E_\pm \pm i \pmag\B_\pm \big) .
\quad
\eea
We can now choose the convenient match
\bea
\Edjm{0B}{d}{njm} &=& -(n+1)\gzBcoef{d}{njm} , 
\nn\\
\Bdjm{1B}{d}{(n-1)jm} 
&=& -\fr{n+1}{d-1}\goBcoef{d}{njm} ,
\nn\\
\Bdjm{1E}{d}{(n-1)jm} 
&=& \fr{n+1}{d-1}\goEcoef{d}{njm}, 
\eea
which gives the first two equations in Eq.\ \rf{SiH}.
The calculation for the CPT-even operators $\Hdual_\eff^\mn$ 
is similar to that for $\gdual_\eff^\mn$, 
up to an overall sign.

\section{Limiting cases}
\label{Limiting cases}

For many applications,
it is appropriate to consider limiting cases
of the Lorentz-violating hamiltonian \rf{pham}.
In this section,
we consider in turn the nonrelativistic limit,
the ultrarelativistic case,
and the restriction to the minimal SME.

\subsection{Nonrelativistic}
\label{Nonrelativistic}

\begin{table*}
\renewcommand{\arraystretch}{1.5}
\begin{tabular*}{0.7\textwidth}{*{4}{@{\extracolsep{\fill}}c}}
Coefficient   & $n$      &  $j$                       & Number \\\hline\hline
$\anr{njm}$   & $\geq 0$ & $n,n-2,n-4,\ldots\geq0$   & $\half(n+1)(n+2)$ \\
$\cnr{njm}$   & $\geq 0$ & $n,n-2,n-4,\ldots\geq0$   & $\half(n+1)(n+2)$ \\
$\gzBnr{njm}$ & $\geq 0$ & $n+1,n-1,n-3,\ldots\geq0$ & $\half(n+2)(n+3)$ \\
$\goBnr{njm}$ & $\geq 0$ & $n+1,n-1,n-3,\ldots\geq1$ & $\half(n+1+\io_n)(n+4-\io_n)$\\
$\goEnr{njm}$ & $\geq 1$ & $n,n-2,n-4,\ldots\geq1$   & $\half(n+1-\io_n)(n+2+\io_n)$\\
$\HzBnr{njm}$ & $\geq 0$ & $n+1,n-1,n-3,\ldots\geq0$ & $\half(n+2)(n+3)$ \\
$\HoBnr{njm}$ & $\geq 0$ & $n+1,n-1,n-3,\ldots\geq1$ & $\half(n+1+\io_n)(n+4-\io_n)$\\
$\HoEnr{njm}$ & $\geq 1$ & $n,n-2,n-4,\ldots\geq1$   & $\half(n+1-\io_n)(n+2+\io_n)$\\
\hline
$\anrfc{n}$  & even, $\geq 0$ & 0 & 1 \\
$\cnrfc{n}$  & even, $\geq 0$ & 0 & 1 \\
$\gnrfc{n}$  & odd, $\geq 0$  & 0 & 1 \\
$\Hnrfc{n}$  & odd, $\geq 0$  & 0 & 1 \\
\hline
\end{tabular*}
\caption{\label{nr_coefs} 
Nonrelativistic coefficients for Lorentz violation.}
\end{table*}

The nonrelativistic limit of the Lorentz-violating hamiltonian \rf{pham}
can be obtained directly from the spherical-harmonic expansions
obtained in Sec.\ \ref{Spherical decomposition}
by expanding the energy $\eomz$ in the usual power series in $\pmag$,  
\beq
\eomz \approx \m + \fr{\pmag^2}{2\m} - \fr{\pmag^4}{8\m^3} +\ldots.
\label{nreomz}
\eeq
In many common physics applications this series can be truncated as desired,
but here it entangles contributions from different dimensions $d$
into any given power $n$ of the momentum $\pmag^n$.
Some care is therefore required in constructing
the nonrelativistic limit.

Consider first $h_a$. 
Substituting the nonrelativistic series \rf{nreomz} for $\eomz$ produces
\bea
h_a &=& \sum_{njm} \pmag^n \, \syjm{0}{jm}(\phat) 
\bigg(\sum_d \m^{d-3-n}
\nn\\
&&
\hskip 20pt
\times \sum_{k\leq n/2} \bc{(d-3-n+2k)/2}{k} \acoef{d}{(n-2k)jm} \bigg),
\qquad 
\label{haexp}
\eea
where $\bc{j}{k}$ denotes a binomial coefficient.
The summation over $k$ represents the 
linear combination of coefficients at dimension $d$ 
appearing in the nonrelativistic limit.
The sum over $d$ gives 
the combination of coefficients of different dimensions 
contributing to the momentum dependence $\pmag^n$.
The expression \rf{haexp} can be viewed an expansion 
in the momentum magnitude $\pmag^n$ and direction $\phat$
involving nonrelativistic coefficients 
consisting of the terms in parentheses.
Each such nonrelativistic coefficient is a superposition 
of the original spherical coefficients 
with fixed values of $j$ and $m$ but summed over $d$ and $k$.
We denote these nonrelativistic coefficients by $\anr{njm}$,
thereby obtaining the nonrelativistic form of $h_a$.
The same reduction can be applied to obtain
the nonrelativistic form of all terms in the hamiltonian \rf{pham}.

The result of this procedure 
is the perturbative nonrelativistic hamiltonian
\bea
\de h^\nr &=& 
h^\nr_a + h^\nr_c 
\nn\\
&&
+ (h^\nr_g)_+ \si^+ + (h^\nr_g)_r \si^r + (h^\nr_g)_- \si^- 
\nn\\
&&
+ (h^\nr_H)_+ \si^+ + (h^\nr_H)_r \si^r + (h^\nr_H)_- \si^- ,
\qquad
\label{phamnr}
\eea
where the spin-independent terms take the form
\bea
h^\nr_a &=& \sum_{njm} \pmag^n
\syjm{0}{jm}(\phat) \anr{njm} ,
\label{ha_nr} 
\nn\\
h^\nr_c &=& -\sum_{njm} \pmag^n
\syjm{0}{jm}(\phat) \cnr{njm} ,
\label{hc_nr}
\eea
and the spin-dependent terms are 
\bea
(h^\nr_g)_r 
&=& - \sum_{njm} \pmag^n \, \syjm{0}{jm}(\phat) \gzBnr{njm} ,
\nn\\
(h^\nr_g)_\pm
&=& \sum_{njm} \pmag^n \, \syjm{\pm 1}{jm}(\phat)
\Big(\pm\goBnr{njm} + i \goEnr{njm} \Big) ,
\nn \\
\nn\\
(h^\nr_H)_r 
&=& \sum_{njm} \pmag^n \, \syjm{0}{jm}(\phat) \HzBnr{njm} ,
\nn \\
(h^\nr_H)_\pm
&=& -\sum_{njm} \pmag^n \, \syjm{\pm 1}{jm}(\phat)
\nn\\
&&
\hskip 40pt
\times
\Big(\pm\HoBnr{njm} + i \HoEnr{njm} \Big) .
\nn\\
\label{SiH_nr}
\eea

The nonrelativistic coefficients 
are related to the spherical coefficients via
\beq
\anr{njm} = \sum_d \m^{d-3-n}
\sum_{k\leq n/2} \bc{(d-3-n+2k)/2}{k} \acoef{d}{(n-2k)jm} ,
\eeq
together with an identical equation relating the coefficients $\cnr{njm}$
to $\ccoef{d}{njm}$, 
and via 
\bea
\gzBnr{njm} &=& \sum_d \m^{d-3-n} 
\hskip-4pt 
\sum_{k\leq n/2} (n-2k+1) 
\nn\\
&&
\hskip 50pt
\times
\bc{(d-4-n+2k)/2}{k} \gzBcoef{d}{(n-2k)jm} ,
\nn\\
\goBnr{njm} &=& \sum_d \m^{d-3-n} 
\hskip-4pt 
\sum_{k\leq n/2} \bc{(d-3-n+2k)/2}{k}
\nn\\
&&
\hskip 30pt
\times
\Big[\goBcoef{d}{(n-2k)jm}+\sqrt{\tfrac{j(j+1)}{2}} 
\gzBcoef{d}{(n-2k)jm} \Big] ,
\nn\\
\goEnr{njm} &=& \sum_d \m^{d-3-n} 
\hskip-4pt 
\sum_{k\leq n/2} \bc{(d-3-n+2k)/2}{k}
\goEcoef{d}{(n-2k)jm} ,
\nn\\
\eea
together with three identical equations relating the coefficients 
$\HzBnr{njm}$ to $\HzBcoef{d}{njm}$,
$\HoBnr{njm}$ to 
$\HoBcoef{d}{njm}$ and $\HzBcoef{d}{njm}$,
and $\HoEnr{njm}$ to $\HoEcoef{d}{njm}$.
In all these expressions,
the entanglement of coefficients with different dimensions $d$
arising from the series \rf{nreomz} is manifest.
For example, 
the coefficients $\acoef{d}{111}$ contribute 
to all $j=m=1$ terms at order $\pmag^n$ for $n=1,3,5,\ldots$.
However, 
for all coefficients
the minimum $d$ required to produce anisotropies 
with a particular $j$ is $j+2$.
This means that probing effects with large $j$ and $n$ 
offers sensitivity to Lorentz violation involving large $d$
that is independent of results from lower values of $n$.

Some properties of the nonrelativistic coefficients
are summarized in Table \ref{nr_coefs}.
The first column lists the coefficients,
while the second column shows the allowed values of $n$.
The third column lists the allowed range of $j$,
while the last column specifies
the number of independent coefficients for each $n$ value.
In this column, $\io_n = 1$ for even $n$
and $\io_n = 0$ for odd $n$.
All nonrelativistic coefficients have mass dimension $1-n$.

In the isotropic limit,
the component nonrelativistic hamiltonians reduce to
\bea
\ring{h}^\nr_a &=& \sum_n \pmag^n \anrfc{n} ,
\nn\\
\ring{h}^\nr_c &=& -\sum_n \pmag^n \cnrfc{n} ,
\nn\\
(\ring{h}^\nr_g)_r &=& -\sum_n \pmag^n \gnrfc{n} ,
\nn\\
(\ring{h}^\nr_H)_r &=& \sum_n \pmag^n \Hnrfc{n} .
\label{SiH_nrfc}
\eea
The nonrelativistic isotropic coefficients in these expressions 
are related to the spherical isotropic coefficients by 
\bea
\anrfc{n} &=& \fr{1}{\sqrt{4\pi}} \anr{n00} ,
\nn\\
\cnrfc{n} &=& \fr{1}{\sqrt{4\pi}} \cnr{n00} ,
\nn\\
\gnrfc{n} &=& \fr{1}{\sqrt{4\pi}} \gzBnr{n00} ,
\nn\\
\Hnrfc{n} &=& \fr{1}{\sqrt{4\pi}} \HzBnr{n00} .
\eea
The nonrelativistic isotropic coefficients have mass dimension $1-n$,
and their index and counting properties are provided in Table \ref{nr_coefs}.
None correspond to Lorentz-invariant operators.

\subsection{Ultrarelativistic}
\label{Ultrarelativistic}

A detailed discussion of the ultrarelativistic limit 
in the context of neutrinos is given in Ref.\ \cite{km12}.
Here,
we consider the single-fermion ultrarelativistic limit 
of the hamiltonian \rf{pham}.
Expanding $\eomz$ gives 
\beq
\eomz \approx 
|\mbf p | + \fr{\m^2}{2|\mbf p|} - \fr{\m^4}{8|\mbf p|^3} +\ldots.
\label{ureomz}
\eeq
However,
substitution of this full series into the spherical decomposition
of the perturbative hamiltonian
generates an expansion in powers of $\pmag^d$ instead of $\pmag^n$,
and the result retains the coefficient complexity 
of the exact expressions \rf{hc} and \rf{SiH}.
For example,
substitution of the full series \rf{ureomz} into the term $h_a$
produces
\beq
h_a = \sum_{djm} \pmag^{d-3}\, \syjm{0}{jm}(\phat)
\sum_{nk} \bc{(d-3-n+2k)/2}{k} \m^{2k} \acoef{d+2k}{njm} ,
\eeq
showing that coefficients with arbitrary $n$ contribute at each $d$.
This situation differs from the nonrelativistic limit,
where substitution of the full analogous series \rf{nreomz}
leads to a simplification of the coefficient structure.
We therefore limit attention here to the dominant term
in the series \rf{ureomz},
which yields the perturbation hamiltonian 
in the ultrarelativistic limit to order $\m$.

To see the effect of taking this ultrarelativistic limit,
consider first $h_a$.
For $\eomz\to\pmag$, 
the above expression reduces to 
\beq
h_a \approx \sum_{djm} \pmag^{d-3} \, \syjm{0}{jm}(\phat) 
\bigg(\sum_n \acoef{d}{njm} \bigg) ,
\eeq
which takes the form of an expansion in $\pmag$ and $\phat$
with ultrarelativistic coefficients 
consisting of the term in parentheses.
We denote these coefficients by $\aur{d}{jm}$.
They are superpositions of spherical coefficients 
with different values of $n$.
Repeating this limiting procedure 
produces the ultrarelativistic limit
of all terms in the hamiltonian \rf{pham}.

The resulting perturbative ultrarelativistic hamiltonian
has the form 
\bea
\de h^\ur &=& h^\ur_a + h^\ur_c 
\nn\\
&&
+ (h^\ur_g)_+ \si^+ + (h^\ur_g)_r \si^r + (h^\ur_g)_- \si^- 
\nn\\
&&
+ (h^\ur_H)_+ \si^+ + (h^\ur_H)_r \si^r + (h^\ur_H)_- \si^- ,
\qquad
\label{phamur}
\eea
where 
the spin-independent terms are 
\bea
h^\ur_a &=& \sum_{djm} \pmag^{d-3}\,
\syjm{0}{jm}(\phat)\, \aur{d}{jm} ,
\label{ha_ur} \\
h^\ur_c &=& -\sum_{djm} \pmag^{d-3}\,
\syjm{0}{jm}(\phat)\, \cur{d}{jm} ,
\label{hc_ur}
\eea
and the spin-dependent terms are 
\bea
(h^\ur_g)_r 
&=& -\m\sum_{djm} \pmag^{d-4}\, \syjm{0}{jm}(\phat) 
\nn\\
&&
\hskip 30pt
\times
\Big[ \gzBur{d}{jm} + \sqrt{\frac{2j}{j+1}} \goBur{d}{jm} \Big] ,
\nn\\
(h^\ur_g)_\pm
&=& \sum_{djm} \pmag^{d-3}\, \, \syjm{\pm 1}{jm}(\phat) 
\nn\\
&&
\hskip 30pt
\times
\Big[\pm\goBur{d}{jm} + i \goEur{d}{jm} \Big] ,
\nn\\
(h^\ur_H)_r 
&=& \m \sum_{djm} \pmag^{d-4}\, \syjm{0}{jm}(\phat) 
\nn\\
&&
\hskip 30pt
\times
\Big[
\HzBur{d}{jm} + \sqrt{\frac{2j}{j+1}} \HoBur{d}{jm} \Big] ,
\nn\\
(h^\ur_H)_\pm
&=& -\sum_{djm} \pmag^{d-3}\, \, \syjm{\pm 1}{jm}(\phat)
\nn\\
&&
\hskip 30pt
\times
\Big[\pm\HoBur{d}{jm} + i \HoEur{d}{jm} \Big] .
\nn\\
\label{SiH_ur}
\eea
Most of the ultrarelativistic coefficients $\Kur{d}{jm}$
are related to the spherical coefficients $\Kcoef{d}{njm}$
by expressions of the form
\beq
\Kur{d}{jm} = \sum_n \Kcoef{d}{njm} .
\eeq
However,
for the $B$-type coefficients it is convenient to define 
\bea
\gzBur{d}{jm} &=& \sum_n \Big[ (n+1-j)\gzBcoef{d}{njm} 
\nn\\
&&
\hskip 50pt
- \sqrt{\frac{2j}{j+1}} \goBcoef{d}{njm} \Big] ,
\nn\\
&&
\nn\\
\goBur{d}{jm} &=& \sum_n \Big[ \goBcoef{d}{njm} 
+ \sqrt{\frac{j(j+1)}{2}} \gzBcoef{d}{njm} \Big] ,
\nn\\
\HzBur{d}{jm} &=& \sum_n \Big[ (n+1-j)\HzBcoef{d}{njm} 
\nn\\
&&
\hskip 50pt
- \sqrt{\frac{2j}{j+1}} \HoBcoef{d}{njm} \Big] ,
\nn\\
&&
\nn\\
\HoBur{d}{jm} &=& \sum_n \Big[ \HoBcoef{d}{njm} 
+ \sqrt{\frac{j(j+1)}{2}} \HzBcoef{d}{njm} \Big].
\qquad
\eea
Each ultrarelativistic coefficient has mass dimension $4-d$.

\begin{table}
\renewcommand{\arraystretch}{1.5}
\begin{tabular*}{0.45\textwidth}{*{4}{@{\extracolsep{\fill}}c}}
Coefficient     & $d$            & $j$                & Number \\
\hline\hline
$\aur{d}{jm}$   & odd, $\geq 3$  & $0\leq j\leq d-2$  & $(d-1)^2$ \\
$\cur{d}{jm}$   & even, $\geq 4$ & $0\leq j\leq d-2$  & $(d-1)^2$ \\
$\gzBur{d}{jm}$ & even, $\geq 4$ & $0\leq j\leq d-3$  & $(d-2)^2$ \\
$\goBur{d}{jm}$ & even, $\geq 4$ & $1\leq j\leq d-2$  & $(d-2)d$ \\
$\goEur{d}{jm}$ & even, $\geq 4$ & $1\leq j\leq d-2$  & $(d-2)d$ \\
$\HzBur{d}{jm}$ & odd, $\geq 5$  & $0\leq j\leq d-3$  & $(d-2)^2$ \\
$\HoBur{d}{jm}$ & odd, $\geq 3$  & $1\leq j\leq d-2$  & $(d-2)d$ \\
$\HoEur{d}{jm}$ & odd, $\geq 3$  & $1\leq j\leq d-2$  & $(d-2)d$ \\
\hline
$\aurfc{d}{}$     & odd, $\geq3$   & 0                  & $1$ \\
$\curfc{d}{}$     & even, $\geq4$  & 0                  & $1$ \\
$\gurfc{d}{}$     & even, $\geq4$  & 0                  & $1$ \\
$\Hurfc{d}{}$     & even, $\geq5$  & 0                  & $1$ \\
\hline
\end{tabular*}
\caption{\label{ur_coefs}
Ultrarelativistic coefficients for Lorentz violation.}
\end{table}

Information about the index ranges and counting 
of the ultrarelativistic coefficients is collected 
in Table \ref{ur_coefs}.
The first column lists the coefficients,
the next two provide the allowed ranges of $d$ and $j$,
and the final column gives the number of independent coefficients
at each $d$.

In the isotropic limit,
the ultrarelativistic hamiltonian components become
\bea
\ring{h}^\ur_a &=& \sum_d \pmag^{d-3} \aurfc{d}{} ,
\nn\\
\ring{h}^\ur_c &=& -\sum_d \pmag^{d-3} \curfc{d}{} ,
\nn\\
(\ring{h}^\ur_g)_r &=& -\m\sum_d \pmag^{d-4} \gurfc{d}{} ,
\nn\\
(\ring{h}^\ur_H)_r &=& \m\sum_d \pmag^{d-4} \Hurfc{d}{} .
\label{SiH_urfc}
\eea
The connection between
the ultrarelativistic isotropic coefficients 
and the spherical isotropic coefficients is 
\bea
\aurfc{d}{} &=& \fr{1}{\sqrt{4\pi}} \aur{d}{00} ,
\nn\\
\curfc{d}{} &=& \fr{1}{\sqrt{4\pi}} \cur{d}{00} ,
\nn\\
\gurfc{d}{} &=& \fr{1}{\sqrt{4\pi}} \gzBur{d}{00} ,
\nn\\
\Hurfc{d}{} &=& \fr{1}{\sqrt{4\pi}} \HzBur{d}{00} .
\eea
The ultrarelativistic isotropic coefficients have mass dimension $4-d$,
and there is no more than one coefficient of any given type at each $d$.
None of them correspond to Lorentz-invariant operators.
The allowed dimensions $d$ are given in Table \ref{ur_coefs}.

We remark in passing that the above results differ in detail
from those obtained in the analysis of the nonminimal neutrino sector
in Ref.\ \cite{km12}.
The differences arise because
the neutrino treatment involves 
Dirac- and Majorana-type couplings 
of multiple flavors of left-handed fermions,
while the present discussion
involves a single Dirac fermion
without helicity restriction.

\subsection{Minimal SME}
\label{Minimal SME}

The minimal SME in flat spacetime 
\cite{ck}
consists of operators of renormalizable dimension $d=3,4$.
In the present context,
this involves the cartesian coefficients
$a^{(3)\mu}$,
$b^{(3)\mu}$,
$c^{(4)\mn}$,
$d^{(4)\mn}$,
$e^{(4)\mu}$,
$f^{(4)\mu}$,
$g^{(4)\la\mn}$,
and $H^{(3)\mn}$.
Of these,
the coefficient $f^{(4)\mu}$ plays no observable role
and can be disregarded
\cite{ba-f,classical},
as described in Sec.\ \ref{Field redefinitions}.
Restricting attention to this coefficient set,
the cartesian expansion  
introduced in Sec.\ \ref{Basics}
can be matched to the spherical-harmonic 
presented in Sec.\ \ref{Spherical decomposition}.
This produces a set of relations connecting 
the minimal cartesian 
and the spherical coefficients for Lorentz violation.

The cartesian expansion of the perturbative hamiltonian \rf{hn}
is given by Eqs.\ \rf{Delta} and \rf{capnmu}.
For the spin-independent piece involving $\De$,
we can use the expressions \rf{effcoeffcomps} 
for the effective coefficients in Eq.\ \rf{aceff}
to project onto the minimal SME terms with $d=3$ and $d=4$,
giving
\bea
\De^{(3)} &=& a^{(3)\mu} p_\mu ,
\nn\\
\De^{(4)} &=& - c^{(4)\mn} p_\mu p_\nu,
\nn\\
\De^{(5)} &=& - \fr {p^2}{\m} e^{(4)\mu} p_\mu.
\label{detff}
\eea
For the spin-dependent terms in $\nv^\mu$,
combining Eq.\ \rf{effcoeffcomps} with Eq.\ \rf{gheff}
and projecting onto the minimal SME coefficients yields
\bea
\nv^{(3)\mu} &=& -\Ht^{(3)\mn}p_\nu ,
\nn\\
\nv^{(4)\mu} &=& \gt_\eff^{(4)\mn\la}p_\nu p_\la 
= \big(
\gt^{(4)\mu\nu\la} - \fr{1}{\m}\et^{\la[\mu}b^{(3)\nu]}
\big) p_\nu p_\la ,
\nn\\
\nv^{(5)\mu} &=& \fr{1}{\m}
( p^\mu d^{(4)\nu\la} p_\nu p_\la -p^2 d^{(4)\mn}p_\nu) .
\label{ntff}
\eea
Note that in these expressions we are using the
dual coefficients $\gt^{(4)\la\mn}$ and $\Ht^{(3)\mn}$
introduced in the expansions \rf{effcompops}.
Since in the minimal case
$\gt^{(4)\mu\nu\la}$ and $b^{(3)\mu}$
always appear in the same linear combination $\gt_\eff^{(4)\mu\nu\la}$,
we use the latter in making the matches that follow.
Note also that the Lorentz-invariant trace 
$d^{(4)\mn}\et_\mn$
is absent from Eq.\ \rf{ntff}.

We can now match these results
to the spherical expansion of the perturbative hamiltonian \rf{hn}
as written in the form \rf{helh}.
This gives relations between the spherical coefficients
and the cartesian ones for the minimal SME.
To express compactly some results,
it is convenient to define an azimuthal spin vector
\beq
\xvec_\pm = \xhat \mp i\yhat .
\eeq

Considering first the match for spin-independent effects,
we find
the four coefficients $a^{(3)\mu}_\eff$ 
are related to the four spherical coefficients $\acoef{3}{njm}$ by
\bea
\acoef{3}{000} &=& \sqrt{4\pi} a^{(3)t}_\eff , 
\nn \\
\acoef{3}{11(-1)} &=& -\sqrt{\fr{2\pi}{3}} x_-^j a^{(3)j}_\eff ,
\nn \\
\acoef{3}{110} &=& -\sqrt{\fr{4\pi}{3}} a^{(3)z}_\eff ,
\nn \\
\acoef{3}{111} &=&  \sqrt{\fr{2\pi}{3}} x_+^j a^{(3)j}_\eff .
\eea
The coefficients $c^{(4)\mn}$ 
are related to the spherical coefficients $\ccoef{4}{njm}$ by
\bea
\ccoef{4}{000} &=& \sqrt{4\pi} c^{(4)tt} ,
\nn \\
\ccoef{4}{11(-1)} &=& -\sqrt{\fr{8\pi}{3}} x_-^j c^{(4)tj} ,
\nn \\
\ccoef{4}{110} &=& -\sqrt{\fr{16\pi}{3}} c^{(4)tz} ,
\nn \\
\ccoef{4}{111} &=& \sqrt{\fr{8\pi}{3}} x_+^j c^{(4)tj} ,
\nn \\
\ccoef{4}{200} &=& \sqrt{\fr{4\pi}{9}} c^{(4)jj} ,
\nn \\
\ccoef{4}{22(-2)} &=& \sqrt{\fr{2\pi}{15}} x_-^j x_-^k c^{(4)jk} ,
\nn \\
\ccoef{4}{22(-1)} &=& \sqrt{\fr{8\pi}{15}} x_-^j  c^{(4)jz} ,
\nn \\
\ccoef{4}{220} &=& 
\sqrt{\fr{4\pi}{5}} \Big( c^{(4)zz} -\frac13 c^{(4)jj} \Big) ,
\nn \\
\ccoef{4}{221} &=& -\sqrt{\fr{8\pi}{15}} x_+^j  c^{(4)jz} ,
\nn \\
\ccoef{4}{222} &=& \sqrt{\fr{2\pi}{15}} x_+^j x_+^k c^{(4)jk} .
\eea
Note that this set of ten coefficients 
contains the trace combination $\et_\mn c^{(4)\mn}$
associated with a Lorentz-invariant operator.
This trace can be removed 
by adding the constraint $\ccoef{4}{000} = 3\ccoef{4}{200}$ 
involving the two isotropic components of $\ccoef{4}{njm}$.
Finally,
as is apparent from Eq.\ \rf{detff},
the minimal cartesian coefficients $e^{(4)\mu}$ act effectively
as $d=5$ coefficients $a^{(5)\mu}$ according to
\bea
\acoef{5}{200} &=& -\acoef{5}{000} 
= \fr{1}{\m} \sqrt{4\pi} e^{(4)t} , 
\nn\\
\acoef{5}{31(-1)} &=& -\acoef{5}{11(-1)} 
= - \fr{1}{\m} \sqrt{\fr{2\pi}{3}} x_-^j e^{(4)j} , 
\nn\\
\acoef{5}{310} &=& -\acoef{5}{110} 
= - \fr{1}{\m} \sqrt{\fr{4\pi}{3}}e^{(4)z} , 
\nn\\
\acoef{5}{311} &=& -\acoef{5}{111} 
= \fr{1}{\m}  \sqrt{\fr{2\pi}{3}} x_+^j e^{(4)j} .
\eea

Turning next to the match for spin-dependent effects,
we begin with the terms involving $\gt_\eff^{(4)\mn\rh}$.
Disregarding the unobservable totally antisymmetric part 
leaves 20 cartesian coefficients,
which can be connected to the twelve $B$-type spherical coefficients 
$\gzBcoef{4}{njm}$, $\goBcoef{4}{njm}$
and the eight $E$-type spherical coefficients $\goEcoef{4}{njm}$.
The nine $B$-type spherical coefficients $\gzBcoef{4}{njm}$ are
related to cartesian ones by 
\bea
\gzBcoef{4}{01(-1)} &=& \sqrt{\fr{2\pi}{3}} x_-^j \gt_\eff^{(4)tjt} ,
\nn \\
\gzBcoef{4}{010} &=& \sqrt{\fr{4\pi}{3}} \gt_\eff^{(4)tzt} ,
\nn \\
\gzBcoef{4}{011} &=& -\sqrt{\fr{2\pi}{3}} x_+^j \gt_\eff^{(4)tjt} ,
\nn \\
\gzBcoef{4}{100} &=& -\sqrt{\fr{\pi}{9}} \gt_\eff^{(4)tjj} ,
\nn \\
\gzBcoef{4}{12(-2)} &=& -\sqrt{\fr{\pi}{30}} x_-^j x_-^k \gt_\eff^{(4)tjk} ,
\nn \\
\gzBcoef{4}{12(-1)} &=& 
-\sqrt{\fr{\pi}{30}} x_-^j  \Big( \gt_\eff^{(4)tjz} + \gt_\eff^{(4)tzj} \Big) ,
\nn \\
\gzBcoef{4}{120} &=& 
-\sqrt{\fr{\pi}{5}} \Big( \gt_\eff^{(4)tzz} - \frac 13 \gt_\eff^{(4)tjj} \Big) ,
\nn \\
\gzBcoef{4}{121} &=& 
\sqrt{\fr{\pi}{30}} x_+^j  \Big( \gt_\eff^{(4)tjz} + \gt_\eff^{(4)tzj} \Big) ,
\nn \\
\gzBcoef{4}{122} &=& -\sqrt{\fr{\pi}{30}} x_+^j x_+^k \gt_\eff^{(4)tjk} .
\eea
The three $B$-type spherical coefficients $\goBcoef{4}{njm}$ 
are given by the equations 
\bea
\goBcoef{4}{21(-1)} &=& -\sqrt{\fr{\pi}{6}} x_-^j \gt_\eff^{(4)jkk} ,
\nn \\
\goBcoef{4}{210} &=& -\sqrt{\fr{\pi}{3}} \gt_\eff^{(4)zjj} ,
\nn \\
\goBcoef{4}{211} &=& \sqrt{\fr{\pi}{6}} x_+^j \gt_\eff^{(4)jkk} ,
\eea
while the expressions for the eight $E$-type spherical coefficients are
\bea
\goEcoef{4}{11(-1)} &=& -i\sqrt{\fr{3\pi}{2}} x_-^j \gt_\eff^{(4)jzt} ,
\nn \\
\goEcoef{4}{110} &=& -i\sqrt{\fr{3\pi}{4}} x_+^j x_-^k \gt_\eff^{(4)jkt} ,
\nn \\
\goEcoef{4}{111} &=& -i\sqrt{\fr{3\pi}{2}} x_+^j \gt_\eff^{(4)jzt} ,
\nn \\
\goEcoef{4}{22(-2)} &=& -i \sqrt{\fr{\pi}{10}} x_-^j x_-^k \gt_\eff^{(4)zjk} ,
\nn \\
\goEcoef{4}{22(-1)} &=& 
i \sqrt{\fr{2\pi}{5}} x_-^j 
\Big( \gt_\eff^{(4)jzz} - \frac 12 \gt_\eff^{(4)jkk} \Big) ,
\nn \\
\goEcoef{4}{220} &=& i \sqrt{\fr{3\pi}{20}} x_+^j x_-^k \gt_\eff^{(4)jkz} ,
\nn \\
\goEcoef{4}{221} &=& 
i \sqrt{\fr{2\pi}{5}} x_+^j 
\Big( \gt_\eff^{(4)jzz} - \frac 12 \gt_\eff^{(4)jkk} \Big) ,
\nn \\
\goEcoef{4}{222} &=& i \sqrt{\fr{\pi}{10}} x_+^j x_+^k \gt_\eff^{(4)zjk} .
\eea
Using the pseudotensor nature of the coefficients $\gt_\eff^{(4)\mn\rh}$, 
one can verify that the $E$-type and $B$-type coefficients 
are associated with operators having parity $(-1)^j$
and $(-1)^{j+1}$, respectively.

The six components of the antisymmetric cartesian coefficients $\Ht^{(3)\mn}$
can be used to obtain 
the three $B$-type spherical coefficients $\HzBcoef{3}{njm}$
and the three $E$-type spherical coefficients $\HoEcoef{3}{njm}$.
The three $B$-type ones are given by 
\bea
\HzBcoef{3}{01(-1)} &=& \sqrt{\fr{2\pi}{3}} x_-^j \Ht^{(3)tj} ,
\nn \\
\HzBcoef{3}{010} &=& \sqrt{\fr{4\pi}{3}} \Ht^{(3)tz} ,
\nn \\
\HzBcoef{3}{011} &=& -\sqrt{\fr{2\pi}{3}} x_+^j \Ht^{(3)tj} ,
\eea
while the three $E$-type ones are found to be 
\bea
\HoEcoef{3}{11(-1)} &=& -i \sqrt{\fr{2\pi}{3}} x_-^j \Ht^{(3)jz} ,
\nn \\
\HoEcoef{3}{110} &=& -i\sqrt{\fr{\pi}{3}} x_+^j x_-^k \Ht^{(3)jk} ,
\nn \\
\HoEcoef{3}{111} &=& -i \sqrt{\fr{2\pi}{3}} x_+^j \Ht^{(3)jz} .
\eea

The remaining 15 cartesian coefficients $d^{(4)\mn}$
specify 15 independent spherical coefficients 
$\HzBcoef{5}{njm}$ and $\HoEcoef{5}{njm}$
corresponding to operators with mass dimension $d=5$,
as can be seen from Eq.\ \rf{ntff}.
The six $B$-type spherical coefficients $\HzBcoef{5}{njm}$ with even $n$
are given by
\bea
\HzBcoef{5}{01(-1)} &=& -\fr{1}{\m}\sqrt{\fr{2\pi}{3}} x_-^j d^{(4)jt} ,
\nn \\
\HzBcoef{5}{010} &=& -\fr{1}{\m}\sqrt{\fr{4\pi}{3}} d^{(4)zt} ,
\nn \\
\HzBcoef{5}{011} &=& \fr{1}{\m}\sqrt{\fr{2\pi}{3}} x_+^j d^{(4)jt} ,
\nn \\
\HzBcoef{5}{21(-1)} &=& -\fr{1}{3\m}\sqrt{\fr{2\pi}{3}} x_-^j d^{(4)tj} ,
\nn \\
\HzBcoef{5}{210} &=& -\fr{1}{3\m}\sqrt{\fr{4\pi}{3}} d^{(4)tz} ,
\nn \\
\HzBcoef{5}{211} &=& \fr{1}{3\m}\sqrt{\fr{2\pi}{3}} x_+^j d^{(4)tj} .
\eea
The six $B$-type spherical coefficients $\HzBcoef{5}{njm}$ with odd $n$
are determined in terms of the six symmetric parts of $d^{(4)jk}$
to be 
\bea
\HzBcoef{5}{100} &=&  \fr{\sqrt\pi}{\m} \Big(d^{(4)tt}+\frac 13d^{(4)jj}\Big) ,
\nn \\
\HzBcoef{5}{12(-2)} &=& \fr{1}{\m}\sqrt{\fr{\pi}{30}} x_-^j x_-^k d^{(4)jk} ,
\nn \\
\HzBcoef{5}{12(-1)} &=& 
\fr{1}{\m}\sqrt{\fr{\pi}{30}} x_-^j \Big(d^{(4)jz}+d^{(4)zj}\Big) ,
\nn \\
\HzBcoef{5}{120} &=& 
\fr{1}{\m}\sqrt{\fr{\pi}{5}} \Big(d^{(4)zz} - \frac 13 d^{(4)jj}\Big) ,
\nn \\
\HzBcoef{5}{121} &=& 
-\fr{1}{\m}\sqrt{\fr{\pi}{30}} x_+^j\Big(d^{(4)jz}+d^{(4)zj}\Big) ,
\nn \\
\HzBcoef{5}{122} &=& \fr{1}{\m}\sqrt{\fr{\pi}{30}} x_+^j x_+^k d^{(4)jk} .
\eea
The three $E$-type spherical coefficients $\HoEcoef{5}{njm}$
are specified in terms of the antisymmetric part of $d^{(4)jk}$ by
\bea
\HoEcoef{5}{11(-1)} &=&
-\fr{i}{\m}\sqrt{\fr{\pi}{6}} x_-^j \Big(d^{(4)jz} - d^{(4)zj}\Big) ,
\nn \\
\HoEcoef{5}{110} &=& 
-\fr{i}{\m}\sqrt{\fr{\pi}{12}} x_+^j x_-^k\Big(d^{(4)jk} - d^{(4)kj}\Big) ,
\nn \\
\HoEcoef{5}{111} &=& 
-\fr{i}{\m}\sqrt{\fr{\pi}{6}} x_+^j \Big(d^{(4)jz} - d^{(4)zj}\Big) .
\eea
Finally,
we remark that the remaining nonzero spherical coefficients 
can be constructed as combinations of the 15 above independent ones
according to
\bea
\HoBcoef{5}{21m} &=& -\HzBcoef{5}{01m}-\HzBcoef{5}{21m} ,
\nn \\
\HoBcoef{5}{32m} &=& -\sqrt{3} \HzBcoef{5}{12m} ,
\nn \\
\HoEcoef{5}{31m} &=& -\HoEcoef{5}{11m} .
\eea
Note that some spherical coefficients remain zero,
as expected from the analysis in Sec.\ \ref{Field redefinitions}
showing that the absorption of the pseudovector operators $\Ahat^\mu$ 
involves only parts of the tensor operators $\That^\mn$.

\section{Applications}
\label{Applications}

Given the spherical decomposition and the various limiting cases,
several immediate applications become feasible. 
In this section,
we begin by revisiting the topics
of dispersion and birefringence
discussed in Sec.\ \ref{Properties},
presenting quantitative expressions 
for the dispersion relation, 
the group velocity,
and the fermion spin precession in various limits.
We next take advantage of the generality of the SME framework 
to make connections to other special models in the literature,
which yields some interesting insights.
With these results in hand,
we can then translate existing astrophysical limits
on isotropic Lorentz violation in the fermion sector 
into constraints on isotropic spherical SME coefficients,
thereby revealing the relationships between the various approaches 
and the scope of the coverage of coefficient space.

\subsection{Dispersion and birefringence}
\label{Dispersion and birefringence}

Using the spherical decomposition
to extend the discussion in Sec.\ \ref{Properties},
we can generate expressions for the dispersion relation,
including several useful limiting cases.
We can also determine the group velocity of a fermion wave packet
and the spin precession of the fermion induced by birefringence.

For simplicity,
we begin by neglecting spin-dependent effects.
Using the result \rf{drapprox} expressed in the spherical basis
and with the birefringent contributions set to zero,
the dispersion relation can be written as
\beq
p^2-\m^2 = 2\eomz \de \eom ,
\label{drsph}
\eeq
where
\beq
\de \eom = \sum_{dnjm}  \eomz^{d-3-n} \pmag^n \syjm{0}{jm} 
\big(\acoef{d}{njm}-\ccoef{d}{njm}\big) .
\eeq
The modified group velocity $\mbf v_g = {\prt \eom}/{\prt\mbf p}$ obeys
\bea
|\mbf v_g|
&=& \fr{\pmag}{\eomz} 
+ \sum_{dnjm} \big((d-3)p^2+ n \m^2\big) \eomz^{d-5-n} \pmag^{n-1}
\nn\\
&&
\hskip 40pt
\times \,
\syjm{0}{jm}(\phat)\, \big(\acoef{d}{njm} - \ccoef{d}{njm}\big).
\eea
Note that either increases or decreases in the velocity are possible,
depending on the signs of the coefficients
and on the direction of travel.
Note also that the corresponding expressions for antifermions
involve opposite signs for the coefficients 
$\acoef{d}{njm}$.

Including spin dependence is straightforward 
in the isotropic limit.
The isotropic dispersion relation also takes the form \rf{drsph}.
Denoting $\eom$ as $\ring\eom$ for this case,
we have 
\bea
\de \ring \eom &=& \sum_{dn} \eomz^{d-3-n} \pmag^n
\big(\afc{d}{n} - \cfc{d}{n} \big)
\nn\\
&&
\pm \m \sum_{dn} \eomz^{d-4-n} \pmag^n 
\big(-\gfc{d}{n} + \Hfc{d}{n} \big)
\nn\\
&=&
\sum_{dn} \eomz^{d-3-n} \pmag^n 
\big( \afc{d}{n} \mp \m \gfc{d+1}{n}
\nn\\
&&
\hskip 80pt
-\cfc{d}{n} \pm \m \Hfc{d+1}{n} \big) .
\qquad
\label{isohel}
\eea
In these expressions, 
the upper and lower signs correspond to
positive and negative helicities, 
respectively.
The modified group velocity in this case is 
\bea
|{\mathaccent'27 {\boldsymbol v}}_g|
&=& \fr{\pmag}{\eomz} 
+ \sum_{dn} \big((d-3)p^2+ n \m^2\big) \eomz^{d-5-n} \pmag^{n-1}
\nn\\
&&
\hskip 30pt
\times
\big(\afc{d}{n} \mp \m \gfc{d+1}{n} 
-\cfc{d}{n} \pm \m \Hfc{d+1}{n} \big) .
\nn\\
\eea
The results for antiparticles
take the same form but with opposite signs for the coefficients
$\afc{d}{n}$ and $\gfc{d+1}{n}$.
We thus see that the two helicities for each fermion species 
and the two for the corresponding antifermions
all experience generically distinct dispersion relations
and group velocities.

Many applications involve fermions at high energies,
where the ultrarelativistic limit may be appropriate.
In this limit,
the dispersion relation takes the form 
\beq
p^2-\m^2 = 2\pmag \de \eom^{\rm UR} ,
\label{drsphur}
\eeq
where
\bea
\de \eom^{\rm UR} &=& 
\sum_{d} \pmag^{d-3} \big(
\aurfc{d}{} \mp \m \gurfc{d+1}{}
\nn\\
&&
\hskip 50pt
-\curfc{d}{} \pm \m \Hurfc{d+1}{} \big) 
\qquad
\eea
in terms of the ultrarelativistic coefficients
defined in Table \ref{ur_coefs}.
The modified group velocity is 
\bea
|\mbf v_g^{\rm UR}|
&=& 1 + \sum_{d} (d-3)\pmag^{d-4}
\big(\aurfc{d}{} \mp \m \gurfc{d+1}{} 
\nn\\
&&
\hskip 50pt
-\curfc{d}{} \pm \m \Hurfc{d+1}{} \big) .
\eea

The above expansions have some intriguing consequences.
One popular approach in the literature focuses
on isotropic modifications to $p^2$ or $\de\ring\eom$
involving powers only of $\pmag$,
restricted to dimensions $d\leq 5$ or $d\leq 6$.
A potentially surprising feature in this context 
is that the expansions of $p^2$ and of $\de\ring\eom$
produce two completely different limits of the general theory. 
As can be seen from Eqs.\ \rf{drsph} and \rf{isohel},
expanding $p^2$ in this way requires imposing the condition $n=d-2$,
while expanding $\de \ring \eom$
requires $n=d-3$ instead,
so the two expansions involve distinct coefficients.
Explicitly,
we find
\bea
p^2-m^2 &=& \mp 2 \m\gfc{4}{1} \pmag
-2\cfc{4}{2} \pmag^2
\nn\\
&&
\mp 2 \m\gfc{6}{3} \pmag^3
-2\cfc{6}{4} \pmag^4 + \ldots
\qquad
\eea
and 
\bea
\de \ring \eom &=& \afc{3}{0} 
\pm \m\Hfc{5}{1} \pmag
\nn\\
&&
+\afc{5}{2} \pmag^2
\pm \m\Hfc{7}{3} \pmag^3
+ \ldots\ ,
\eea
showing that the two approaches have orthogonal content.
Note that both expansions contain terms
with odd and even powers of $\pmag$,
but the first involves only operators of even dimension $d$
while the second involves only operators of odd $d$.
The attribution of operator dimensionality in this way
is an automatic and natural consequence
of the freedom to use field redefinitions
to absorb some effects into others,
discussed in Sec.\ \ref{Field redefinitions}.

The isotropic expansions of $p^2$ and $\de\ring\eom$
can be arranged to match
if we stipulate {\it a priori} 
that only ultrarelativistic physics is relevant. 
In the ultrarelavistic limit,
we obtain 
\bea
p^2-m^2 &=& 
2\big(\aurfc{3}{} \mp \m \gurfc{4}{} \big) \pmag 
\nn\\
&&
+ 2\big(-\curfc{4}{} \pm \m \Hurfc{5}{} \big) \pmag^2
\nn\\
&&
+ 2\big(\aurfc{5}{} \mp \m \gurfc{6}{} \big) \pmag^3 
\nn\\
&&
+ 2\big(-\curfc{6}{} \pm \m \Hurfc{7}{} \big) \pmag^4
+ \ldots
\nn\\
&=& 2 \pmag\de\eom^{\rm UR}.
\label{urdr}
\eea
This expression reveals
that the natural attribution of operator dimensionalities
in the ultrarelavistic expansions
involves a mixing of operators of different $d$
at each power of $\pmag$. 

The components $h_g$ and $h_H$
of the perturbative hamiltonian
give rise to birefringence,
which can be viewed as a Larmor-like precession of the spin $\Svec$
as the particle travels.
Writing the expressions \rf{swc} in the form
$h_g = \mbf{h}_g\cdot \mbf\si$ and $h_H = \mbf{h}_H\cdot \mbf\si$,
the rate of change of the spin expectation value 
of a particle state localized in momentum space is
given via the commutator of the hamiltonian $h$ with the spin $\Svec$ as 
\bea
\fr{d\vev{\Svec}}{dt} &=& \vev{i[h,\Svec]}
\approx 2(\mbf{h}_g+\mbf{h}_H) \times\vev{\Svec} .
\label{sppr}
\eea
The precession frequency is then
$\mbf{\om} = 2(\mbf{h}_g + \mbf{h}_H)$.
This generalizes the result obtained for muon precession
and used to extract constraints on muon Lorentz violation 
from storage-ring data
\cite{muexpt}.
In the helicity basis, 
we can write the result \rf{sppr} in component form as
\beq
\fr{d\vev{S^u}}{dt} = \ep^{uvw} 2(h_g+h_h)_v \vev{S_w} ,
\eeq
where $u$, $v$, $w$ range over components labeled by $(+,r,-)$
and the nonzero components of the antisymmetric tensor $\ep^{uvw}$
are specified by $\ep^{+r-} = -i$.

In the isotropic limit, 
the nonzero spin-dependent terms are given by $(h_g)_r$ and $(h_H)_r$.
The helicity states then become stationary states,
and the expression for the spin precession takes the simple form
\beq
\fr{d\vev{S^\pm}}{dt} = \mp2i (h_g + h_H)_r \vev{S^\pm} , 
\eeq
where
\beq
(h_g + h_H)_r = -\m\sum_{dn} \eomz^{d-4-n} \pmag^n (\gfc{d}{n}-\Hfc{d}{n}) .
\eeq
The expectation value $\vev{S^r}$ of the helicity 
remains constant in this limit.

\subsection{Connections to other formalisms}
\label{Connections to other formalisms}

A few special models containing quadratic fermion operators with $d>4$
can be found in the existing literature. 
The generality of the SME-based analysis presented above
implies that any special model based on standard field theory
can be described using a selected subset 
of the cartesian coefficients in Table \ref{free_summary}
or, equivalently,
of the spherical coefficients in Table \ref{spherical_coefs}.
The SME framework also incorporates
several kinematical frameworks in a field-theoretic context.
In this subsection,
we summarize some of these links,
treating first field-theoretic models and then kinematical formalisms.

\subsubsection{Field-theoretic models}
\label{Field-theoretic models}

Consider first special models 
defined via a Lagrange density for a Dirac fermion of mass $\m$
that contains quadratic fermion operators with $d>4$.
Examples in the literature include models 
with a few specific Lorentz-violating operators
of dimensions $d=5$ and $d=6$. 
Here,
we identify the correspondence between these models
and the SME coefficients for Lorentz violation. 

One special model with quadratic Dirac operators
is given by Myers and Pospelov 
\cite{rmmp}.
This model involves two $d=5$ operators for Lorentz violation 
constructed using a timelike vector $n^\mu$,
which fixes a preferred frame,
and two corresponding parameters 
$\et_1/M_P$, $\et_2/M_P$.
Matching the operators to the SME framework
reveals that the nonzero cartesian coefficients
for Lorentz violation are given by 
\beq
a^{(5)\mu\al\be} = \fr{\et_1}{M_P} n^\mu n^\al n^\be , 
\quad
b^{(5)\mu\al\be} = -\fr{\et_2}{M_P} n^\mu n^\al n^\be .
\label{abrestrict}
\eeq
In the preferred frame with $n^\mu = (1,0,0,0)$,
the model is isotropic 
and can therefore be matched to isotropic spherical coefficients
in the SME.
We find the correspondence
\bea
\afc{5}{0} = \fr{\et_1}{M_P} , 
\quad
\gfc{6}{1} = -\fr{\et_2}{\m M_P} .
\eea
The model therefore involves two 
of the eight possible observable isotropic degrees of freedom
with $d=5$ and $d=6$ listed in Table \ref{spherical_coefs}
and displayed explicitly in Eqs.\ \rf{five} and \rf{six}:
one of the three for $d=5$,
and one of the five for $d=6$.
Note that the parameter $\et_2/M_P$
is most naturally viewed as an observable isotropic $d=6$ coefficient
due to the freedom to make field redefinitions
absorbing all $\bhat^\mu$ coefficients,
as discussed in Sec.\ \ref{Field redefinitions}.

An extension of this model is given by Mattingly 
\cite{dm},
who uses the notation 
$u^\al\equiv n^\al$, 
$E_P\equiv M_P$.
In addition to the two operators \rf{abrestrict},
this extension includes two others with $d=5$
parametrized by 
$\al^{(5)}_L/M_P$, $\al^{(5)}_R/M_P$,
and four more with $d=6$
controlled by the real parameters
$\al^{(6)}_L/M_P$, $\al^{(6)}_R/M_P$,
$\widetilde\al^{(6)}_L/M_P$, $\widetilde\al^{(6)}_R/M_P$.
Matching the $d=5$ terms 
to the cartesian coefficients in the SME 
yields nonzero contributions 
\bea
m^{(5)\al\be} &=& -\fr{(\al^{(5)}_L+\al^{(5)}_R)}{2M_P}n^\al n^\be ,
\nn\\
m_5^{(5)\al\be} &=& -i\fr{(\al^{(5)}_L-\al^{(5)}_R)}{2M_P}n^\al n^\be .
\label{mlike}
\eea
In the SME framework,
hermiticity requires $\al^{(5)}_R = \al^{(5)*}_L$.
This condition appears to have been overlooked in the literature.
If $\al^{(5)}_L$ and $\al^{(5)}_R$ are both real
then only $m^{(5)\al\be}$ is nonzero;
if both parameters are imaginary,
then only $m_5^{(5)\al\be}$ is nonzero;
while even when both parameters are complex 
only two degrees of freedom appear.
For the $d=6$ terms,
the corresponding nonzero cartesian coefficients in the SME 
are given by 
\bea
c^{(6)\mu\al\be\ga} &=& 
\fr{(\al_L^{(6)} + \al_R^{(6)})}{2M_P^2}
n^\mu n^\al n^\be n^\ga
\nn\\
&&
+ \fr{(\widetilde\al_L^{(6)} + \widetilde\al_R^{(6)})}{2M_P^2}
n^\mu n^\al \et^{\be\ga} ,
\nn\\
d^{(6)\mu\al\be\ga} &=&
\fr{(\al_L^{(6)} - \al_R^{(6)})}{2M_P^2}
n^\mu n^\al n^\be n^\ga
\nn\\
&&
+\fr{(\widetilde\al_L^{(6)} - \widetilde\al_R^{(6)})}{2M_P^2}
n^\mu n^\al \et^{\be\ga} .
\eea
This model is also isotropic in the preferred frame with $n^\mu = (1,0,0,0)$.
Matching all the additional terms
to the isotropic spherical coefficients in the SME
in this frame gives
\bea
\cfc{6}{0} &=& 
\fr{\al_L^{(5)}+\al_R^{(5)}}{2\m M_P} 
+\fr{\al_L^{(6)}+\al_R^{(6)} 
+\widetilde\al_L^{(6)}+\widetilde\al_R^{(6)}}{2M_P^2},
\nn\\
\cfc{6}{2} &=& 
-\fr{\al_L^{(5)}+\al_R^{(5)}}{2\m M_P} 
-\fr{\widetilde\al_L^{(6)}+\widetilde\al_R^{(6)}}{2M_P^2} ,
\nn\\
\Hfc{7}{1} &=& \fr{\al_L^{(6)}-\al_R^{(6)}
  +\widetilde\al_L^{(6)}-\widetilde\al_R^{(6)}}{2\m M_P^2} ,
\nn\\
\Hfc{7}{3} &=& -\fr{\widetilde\al_L^{(6)}-\widetilde\al_R^{(6)}}{2\m M_P^2} .
\eea
This match reveals that the couplings 
$\al^{(5)}_L$, $\al^{(5)}_R$ 
are most naturally viewed as a single real observable isotropic coupling
at $d=6$,
involving observable effects that are inseparable from those governed by 
the coefficient sum $\widetilde\al^{(6)}_L+\widetilde\al^{(6)}_R$.
The combination associated with $m_5^{(5)\al\be}$
in Eq.\ \rf{mlike} has no observable effects,
as shown in Sec.\ \ref{Field redefinitions}.
Also,
the four degrees of freedom in the real parameters 
$\al^{(6)}_L$, $\al^{(6)}_R$,
$\widetilde\al^{(6)}_L$, $\widetilde\al^{(6)}_R$
are most naturally interpreted as 
two of the five observable isotropic coefficients with $d=6$ 
and two of the five with $d=7$.

Another special model involving quadratic Dirac operators
is considered by Rubtsov, Satunin, and Sibiryakov 
\cite{rss}.
This isotropic model contains a $d=4$ term and a $d=6$ term,
with parameters $\vka$ and $g$.
It corresponds to the SME framework in the limit
\beq
c^{(4)jk} = -\vka \de^{jk} , 
\quad 
c^{(6)jklm} = -g \de^{jk}\de^{lm} .
\eeq
In the preferred frame,
the match to 
the isotropic spherical coefficients in the SME is 
\beq
\cfc{4}{2} = -\vka, 
\quad
\cfc{6}{4} = -g/M^2,
\eeq
showing that the model involves 
another of the five possible isotropic operators for $d=6$
displayed in Table \ref{spherical_coefs} and Eq.\ \rf{six}.

A more general model focusing on $d=5$ operators
is given by Bolokhov and Pospelov 
\cite{pbmp},
who limit attention to operators that
cannot be reduced to ones with $d<5$ using the equations of motion.
The model involves quadratic fermion operators
expressed in terms of parameters
$h_1^{\al\be}$, 
$h_2^{\al\be}$, 
$C_1^{\mu\al\be}$,
$C_2^{\mu\al\be}$,
$E_1^{\mu\nu\al\be}$
and $E_4^{(d)\al\mu\nu\be}$.
These parameters form a subset of the $d=5$ cartesian coefficients 
listed in Table \ref{free_summary}.
Explicitly, 
we find the relations 
\bea
m^{(5)\al\be} &=& 2 h_1^{\al\be} , 
\quad
m_5^{(5)\al\be} = -2i h_2^{\al\be} ,
\nn\\
a^{(5)\mu\al\be} &=& 6 C_1^{\mu\al\be} ,
\quad
b^{(5)\mu\al\be} = -6 C_2^{\mu\al\be} ,
\nn\\
H^{(5)\mu\nu\al\be} &=& 12 E_1^{\mu\nu\al\be} + 16 E_4^{(d)\al\mu\nu\be} .
\eea
Counting the degrees of freedom in each relation is instructive. 
The parameters $h_1^{\al\be}$ and $h_2^{\al\be}$
are symmetric and so each have 10 independent components,
matching the SME counting for $m^{(5)\al\be}$ and $m_5^{(5)\al\be}$.
The parameters $C_1^{\mu\al\be}$ and $C_2^{\mu\al\be}$
are totally symmetric,
giving $20+20$ independent components,
whereas the SME coefficients
$a^{(5)\mu\al\be}$ and $b^{(5)\mu\al\be}$
contain a total of $40+40$ degrees of freedom.
We remark in passing that the 20 parameters $C_1^{\mu\nu\rh}$ 
correspond to the 20 on-shell effective coefficients 
$a_\eff^{(5)\al\be\ga}$
in Eq.\ \rf{aceff}.
The parameter $E_1^{\mu\nu\al\be}$ 
is antisymmetrized in $\mu\nu$ 
and then symmetrized in $\mu\nu\be$,
giving a total of 45 independent components,
while $E_4^{\al\mu\nu\be}$ 
is antisymmetrized in $\al\mu\nu$ 
and then symmetrized in $\al\be$,
generating 15 independent components.
The total is $45+15=60$, 
matching the counting for the SME
coefficients $H^{(5)\mu\nu\al\be}$.
Note,
however,
that the 45 components of $E_1^{\mu\nu\al\be}$
cannot be matched to the 45 on-shell SME effective coefficients
$H_\eff^{(5)\mu\al\be\ga}$ in Eq.\ \rf{gheff}.

\subsubsection{Kinematical formalisms}
\label{Kinematical formalisms}

In the context of the photon sector,
Sec.\ IV F of Ref.\ \cite{km09}
discusses the relationship between the SME
and several kinematical formalisms 
purporting to describe aspects of Lorentz violation
based on modifications of the transformation laws.
In this subsection,
we revisit these discussions briefly
in light of the insights provided by the nonminimal fermion sector.

One kinematical approach involves models called  
deformed special relativities (DSR). 
These are defined as smooth nonlinear momentum-space representations 
of the usual Lorentz transformations,
which is known to imply that they have no observable consequences 
beyond conventional special relativity
\cite{dsr}.
All corresponding coefficients for Lorentz violation
in the photon sector are explicitly constructed in
in Sec.\ IV F 3 of Ref.\ \cite{km09}
and are indeed found to be unobservable.
Since DSR models are sector-independent by definition,
a parallel analysis holds for the nonminimal fermion sector
discussed in the present work,
and so further consideration of DSR models in this context 
provides no new insights.

Another kinematical approach,
the Robertson-Mansouri-Sexl (RMS) formalism  
\cite{rms},
does describe certain physical deviations from special relativity.
The RMS formalism can be viewed as a special limit of the SME 
requiring flat spacetime, 
the existence of a universal preferred frame U
in which light is conventional,
and only isotropic Lorentz violation affecting clocks and rods in U.
The three RMS parameters $a$, $b$, $d$ are experiment dependent
unless identical clocks and rods in the same physical states are used
as the reference standards,
so caution is required in comparing results from different experiments.
For any given experiment,
the RMS parameters can in principle be expressed in terms of SME coefficients
by incorporating the underlying physics of the clocks and rods.
The mapping from the SME to the RMS formalism is described 
in Sec.\ IV F 2 of Ref.\ \cite{km09}.

The development of the nonminimal fermion sector in the present work 
offers the opportunity to investigate further
the relationship between the RMS formalism and the SME
by considering effects from the fermion content of clocks and rods.
In general, 
the behavior of physical clocks and rods is complicated 
and determined by properties of their component particles 
and the forces involved.
A detailed SME description is therefore necessary 
for a careful treatment of Lorentz violation in this context. 
However,
a phenomenological treatment in the SME vein
using simple model clocks and rods
can illustrate some of the basic features 
to be expected from Lorentz violation
and their role in the RMS formalism.

Consider first a clock in conventional special relativity 
that ticks at a frequency $\om_0$ in a comoving inertial frame.
In a different boosted frame,
the frequency $\om\equiv p^0$ of the clock 
and the wave vector $\pvec$ of its oscillations
obey a dispersion-type relation
$p_\mu p^\mu \equiv \om^2 - \pvec^2 = \om_0^2$,
where the invariant $\om_0$ plays the role of a particle mass.
In the presence of Lorentz violation,
this dispersion relation becomes modified.
Ignoring possible spin effects for simplicity,
the modified relation can be written
\beq
p^2 = \om_0^2 + 2\ahat_c - 2\chat_c,
\label{mdrc}
\eeq
where $\ahat_c$ and $\chat_c$ 
are $p^\mu$-dependent effective Lorentz-violating corrections 
associated with CPT-odd and CPT-even operators, respectively. 
If the clock is a single fermion of mass $\m$,
Eq.\ \rf{mdrc}
can be viewed as a special limit 
of the modified dispersion relation \rf{drapprox}
derived in Sec.\ \ref{Properties}
\cite{massclock}.
To match to the RMS formalism,
we must further restrict the clock dispersion relation
by assuming the existence of a preferred universal frame U
in which the physics describing the clock is isotropic.
The dispersion relation \rf{mdrc} then takes the form
\beq
p^2 = \om_0^2 
+ 2\sum_{dn} \om^{d-2-n} \pmag^n \Big(\aclock{d}{n} - \cclock{d}{n}\Big),
\label{clockdr}
\eeq
involving only the isotropic effective coefficients for Lorentz violation
$\aclock{d}{n}$ and $\cclock{d}{n}$.
The allowed values of $d$ and $n$ 
and the coefficient counting 
are the same as those given in the last four rows 
of Table \ref{spherical_coefs}.

Suppose a clock obeying \rf{clockdr} moves at a constant speed $v$ 
in the $x$ direction relative to U.
In the comoving inertial frame,
the clock wave 4-vector can be written as 
$k^\mu = (\om_c,0,0,0)$,
where $\om_c$ denotes the ticking frequency.
In the frame U,
the wave 4-vector takes the form $p^\mu = (\ga\om_c,\ga v\om_c,0,0)$.
Combining this expression with the dispersion relation \rf{clockdr},
the velocity-dependent ratio 
of the clock ticking frequencies $\om_c(v)$ and $\om_c(0)\equiv\om_c$ is 
\beq
\fr{\om_c(v)}{\om_c(0)}
= 1 + \sum_{dn} \om_0^{d-4} \Big(v^n\ga^{d-2}-\de_{n0}\Big) 
\Big(\aclock{d}{n}-\cclock{d}{n}\Big) .
\label{cratio}
\eeq
In the SME, 
the frequency $\om_c(v)$ is frame dependent.
In the RMS formalism,
however,
this clock serves as the time standard,
with all other times measured relative to it.
The ratio \rf{cratio} reduces to $1$ for vanishing $v$
and is an even function of $v$ because $n$ is even,
in agreement with RMS postulates.

Next,
consider a rod in conventional special relativity
whose length and orientation
are specified by a rest-frame wave vector $\kvec_0$
and a corresponding wave 4-vector $k^\mu = (0;\kvec_0)$.
A simple choice of rod is the Compton wavelength of a single particle,
which could be of a species different 
from any involved in the time standard.
Other rod choices are possible,
such as one formed from particles 
with frequencies locked to an internal clock.
In a boosted frame,
the wave 4-vector $p^\mu$ of the rod obeys the dispersion-type relation
$p^2 = -\komag^2$. 
In the presence of Lorentz violation,
and assuming as before the existence of a universal preferred frame U
as required by the RMS formalism,
the modified dispersion relation for the rod can be written as 
\beq
p^2 = -\komag^2 
+ 2\sum_{dn} \om^{d-2-n} \pmag^n \Big(\arod{d}{n} - \crod{d}{n}\Big) ,
\label{roddr}
\eeq
where the allowed values of $d$ and $n$ and the coefficient counting
parallel those for the clock. 

If a rod with wave vector $\kvec_r$ in a comoving frame
moves at speed $v$ in the $x$ direction relative to the frame U,
then its wave 4-vector in U is 
$p^\mu = (\ga v k_r^x, \ga k_r^x, k_r^y, k_r^z)$.
Using the dispersion relation \rf{roddr}
reveals that the velocity-dependent ratio
of the wave-vector magnitudes
$\krmag \vvec$ and $\krmag 0 \equiv |\mbf k_r|$
is 
\bea
\fr{\krmag \vvec }{\krmag 0} &=& 1 
\nn\\
&&
\hskip -30pt
- \sum_{dn} \komag^{d-4} 
\Big(\ga^{d-2}v_\parallel^{d-2-n}(1-v_\perp^2)^{n/2} - \de_{n,d-2}\Big)
\nn\\
&&
\hskip 40pt
\times \Big(\arod{d}{n} - \crod{d}{n}\Big) ,
\label{rratio}
\eea
where $v_\parallel$ and $v_\perp$ are the components
of the boost velocity parallel and perpendicular to the rod,
respectively.
This expression characterizes the variations in rod length
in different Lorentz frames,
explicitly showing the orientation and velocity dependence 
arising in the SME context.
In contrast,
the rod serves as the length standard in the RMS formalism, 
with all other lengths measured relative to it.

The result \rf{rratio} illuminates some aspects of the RMS formalism.
The coefficients $\crod{d}{n}$ associated with CPT-even effects
have indices $d$ and $n$ taking only even values,
and hence they introduce dilations involving only even powers of $v$.
This is in agreement with the RMS postulates.
However, 
the coefficients $\arod{d}{n}$ controlling CPT-odd effects
produce shifts that are odd in $v$, 
so boosts in opposite directions give different effects.
This possibility lies outside the RMS formalism
despite its origin in comparatively simple 
isotropic Lorentz violations in U.

The expression \rf{rratio} has another significant implication:
the rod length measured in RMS coordinates 
is the same when the rod is oriented along any of the three coordinate axes,
but it typically differs for other orientations.
This feature appears to have been overlooked in the literature.
It emerges here in the context of a simple SME-based model,
but the dependence of the ratio \rf{rratio} 
on parallel and perpendicular velocities
suggests it is a generic aspect of Lorentz violation.
In particular,
the RMS transformation assumes one rod is aligned
along the boost axis and the other two are perpendicular. 
The expression \rf{rratio} therefore implies 
that nonstandard choices of rod orientation
lie outside the RMS formalism
because they cannot be linked to the frame U
via a transformation of the RMS form.
This is problematic for laboratory experiments
attempting to report bounds in the RMS language
because the results of any measurement
are meaningful only when the chosen length standards 
are correctly aligned with a particular boost
and moreover only when this alignment is maintained 
throughout the measurement process.
This requirement is challenging and perhaps impossible
to satisfy in practice 
due to the rotation and orbital revolution of the Earth
and to the motion of the solar system relative to the frame U.

For the simple model 
with clocks of type \rf{clockdr} and rods of type \rf{roddr}
with RMS-compatible orientations,
we can construct explicitly 
the RMS transformation $T$ from U to the boosted frame
and identify the RMS parameters $a$, $b$, $d$
and hence the factors $\al$, $\be$, and $\de$
multiplying their $v^2$ components. 
Assuming Einstein synchronization,
$T$ takes the form 
\beq
T = \begin{pmatrix}
  a\ga^2 & -av\ga^2 & 0 & 0 \\
  -bv & b & 0 & 0 \\
  0 & 0 & d & 0 \\
  0 & 0 & 0 & d
\end{pmatrix} ,
\eeq
where $a$, $b$, and $d$ are functions of $v$
that reduce to $a=1/\ga$, $b=\ga$, and $d=1$
in the Lorentz-invariant limit.
The RMS transformation can be viewed as the product $T=C\La$
of a standard Lorentz transformation $\La$ 
from U to the comoving Lorentz frame 
with coordinate dilations $C$ scaling space and time 
relative to the chosen clocks and rods 
\cite{km09}.
In terms of RMS functions,
the scaling matrix $C$ is diagonal with entries
$(a\ga,b/\ga,d,d)$.
The ratio \rf{cratio} then implies 
that the time-dilation function $a$ is given by
\bea
a &=& \fr 1 \ga 
+ \fr 1 \ga \sum_{dn} \om_0^{d-4} \Big(v^n\ga^{d-2}-\de_{n0}\Big) 
\Big(\aclock{d}{n}-\cclock{d}{n}\Big) 
\nn\\
&=& 1 + v^2 \Big[-\half + \sum_d \om_0^{d-4}
\Big( \frac{d-2}{2}\aclock{d}{0} - \frac{d-2}{2}\cclock{d}{0}
\nn\\
&&
\hskip 80 pt
+\aclock{d}{2} - \cclock{d}{2}\Big)\Big] 
+ O(v^4).
\nn\\
\eea
The coefficient of $v^2$ is 
the expression for the RMS parameter $\al$
in terms of SME coefficients for Lorentz violation
in this simple model.

To find the RMS functions $b$ and $d$ for spatial dilations,
consider first a rod oriented along the boost direction $x$ 
and two rods in the orthogonal $y$ and $z$ directions.
For the rod lying along the $x$ axis,
the ratio \rf{rratio} gives 
\bea
b &=& \ga  - \ga \sum_{dn} \komag^{d-4}
\Big(\ga^{d-2}v^{d-2-n} - \de_{n,d-2}\Big)
\nn\\
&&
\hskip 80 pt
\times
\Big(\arod{d}{n} - \crod{d}{n}\Big) 
\nn\\
&=& 1 - v \sum_d \komag^{d-4} \arod{d}{d-3}
\nn\\
&&
\hskip 10 pt
+ v^2\Big[\half +\sum_d \komag^{d-4} 
\Big(\crod{d}{d-4}+\tfrac{d-2}{2} \crod{d}{d-2}\Big) \Big] 
\nn\\
&&
\hskip 10 pt
+ O(v^3) .
\eea
The term linear in the boost $v$ 
stems from CPT violation and lies outside the RMS formalism,
as discussed above.
The coefficient of the term quadratic in $v^2$ 
is the RMS parameter $\be$.

\begin{table*}
\renewcommand{\arraystretch}{1.2}
\setlength{\tabcolsep}{5pt}
\begin{tabular}{c|c|c|c|c|c}
Dimension & Sector & Lower bound & Coefficient & Upper bound & Source \\
\hline
\hline
&&&&&\\[-10pt]
$d=4$ & electron &
& $\curfc{4}{e}$ & $< 1.5\times10^{-15}$ & \cite{cg1} \\
&&$-5\times10^{-13} <$ & $\curfc{4}{e}$ & & \cite{cg2} \\
&&$-1.3\times10^{-15} <$ & $\curfc{4}{e}$ & $< 2\times10^{-16}$ & \cite{gs} \\
&&$-1.2\times10^{-16} < $ & $\curfc{4}{e}$ & & \cite{jlm1} \\
&&$-6\times10^{-20} < $ & $\curfc{4}{e}$ & & \cite{fs13} \\
& proton &
$-5\times10^{-23} <$ & $\curfc{4}{p}$ & & \cite{cg1} \\
&&& $\curfc{4}{p}$ & $< 5\times10^{-24}$ & \cite{cg2} \\
&&$-2\times10^{-22} <$ & $\curfc{4}{p}$ & & \cite{jlm1} \\
&&& $\curfc{4}{p}$ & $< 4.5\times10^{-23}$ & \cite{ss}\\
&&$-9.8\times10^{-22} <$ & $\curfc{4}{p}-\curfc{4}{e}$ & $< 9.8\times10^{-22}$ & \cite{ap} \\
& quark &
$-1\times10^{-23} <$ & $\curfc{4}{q}$ & $<1.8\times10^{-21}$ & \cite{gm} \\
&&$-1\times10^{-23} <$ & $\curfc{4}{q}-2\curfc{4}{e}$ & $< 2\times10^{-20}$ & \cite{gm} \\
[2pt]\hline &&&&&\\[-10pt]
$d=5$ & electron &
&$\aurfc{5}{e}$ & $< 6.5\times10^{-27}$ & \cite{konmaj} \\
&&$-3.5\times10^{-27} <$ & $\aurfc{5}{e}$ & & \cite{jlm2} \\
&&$-1\times10^{-34} <$ & $\aurfc{5}{e}-m_e \gurfc{6}{e}$ & $< 1\times10^{-34}$ & \cite{gm} \\
&&$-4\times10^{-25} <$ & $\aurfc{5}{e} \pm m_e\gurfc{6}{e}$ & $< 4\times10^{-25}$ & \cite{ml1} \\
&&$-1\times10^{-20} <$ & $\aurfc{5}{e}$ & $< 2.8\times10^{-17}$ & \cite{ba11} \\
& muon &
$-1\times10^{-34} <$ & $\aurfc{5}{\mu}-m_\mu \gurfc{6}{\mu}$ & $< 1\times10^{-34}$ & \cite{gm} \\
& tau &
$-2\times10^{-33} <$ & $\aurfc{5}{\ta}-m_\ta\gurfc{6}{\ta}$ & $< 2\times10^{-33}$ & \cite{gm} \\
[2pt]\hline &&&&&\\[-10pt]
$d=6$ & electron &
$-8.5\times10^{-20} <$ & $\curfc{6}{e}$ & $< 2.5\times10^{-23}$ & \cite{gm} \\
&&$-5.4\times10^{-14} <$ & $\gurfc{6}{e}$ & $< 5.4\times10^{-14}$ & \cite{ba11} \\
& muon &
$-8.5\times10^{-20} <$ & $\curfc{6}{\mu}$ & $< 2.5\times10^{-23}$ & \cite{gm} \\
& proton &
$-3.4\times10^{-45} <$ & $\curfc{6}{p}$ & $< 3.4\times10^{-42}$ & \cite{ml2} \\
& quark &
$-6.3\times10^{-23} <$ & $\curfc{6}{q}$ & $< 1.7\times10^{-22}$ & \cite{gm} \\
\hline
\end{tabular}
\caption{
Astrophysical limits on isotropic SME coefficients.
Units are GeV$^{4-d}$.}
\label{tabletranslation}
\end{table*}

Next, consider a rod lying along the $y$ or $z$ axis.
In the simple model with the ratio \rf{rratio},
no dilation along these directions is produced
and so the RMS parameter $d$ is found to be $d=1$,
implying $\de = 0$.
However, 
the result \rf{rratio}
accounts only for modifications arising 
from the coupling of the intrinsic wavelengths of the rod components 
to the Lorentz-violating vacuum.
A more realistic phenomenological description of a rod
must also allow for couplings of the rod bulk properties
such as its macroscopic momentum or spin.
For example,
if the rod has mass $M$,
then its bulk 4-momentum
in the boosted frame takes the form $P^\mu = M\ga(1,\vvec)$.
Suppose the effective dispersion relation for the rod can be written as 
\beq
p^2 = -\komag^2 - 2C_r \Pvec^2 
\eeq
instead of the result \rf{roddr}.
The modification vanishes when the rod is at rest in U,
but otherwise leads to an isotropic rod distortion given by
\bea
\fr{b}{\ga} = d 
&=& \fr{\krmag v}{\krmag 0} 
= 1 + \komag^{-2}C_r M^2 \ga^2 v^2 
\nn\\
&=& 1 + (\komag^{-2}C_r M^2) v^2 + O(v^4) .
\qquad 
\label{momdr}
\eea
In this case,
the RMS parameter $d$ is nonzero
and the coefficient multiplying $v^2$
is the parameter $\de = \be - \half$.
Note that in more realistic models the parameters $\be$ and $\de$
are independent and both nonzero.
For example,
the simple phenomenological model
obtained by adding the two modifications \rf{roddr} and \rf{momdr}
generates independent nonzero parameters $\be$ and $\de$.

\subsection{Astrophysical bounds}
\label{Astrophysical bounds}

A number of papers in the literature
obtain bounds on various kinds of isotropic Lorentz violation
from astrophysical observations.
A few of these are based on field-theoretic models,
but the bulk use an approach based on isotropic dispersion relations. 
The results obtained in Secs.\ \ref{Dispersion and birefringence}
and \ref{Field-theoretic models}
make feasible a translation of these various bounds 
into constraints on isotropic spherical coefficients in the SME.
This translation also clarifies the relationships
between the different bounds 
and reveals the coverage of the available coefficient space
achieved to date.

Since all the astrophysical bounds are obtained at high energies,
it is appropriate to work in the ultrarelativistic limit of the SME,
with dispersion relation given by Eq.\ \rf{urdr}.
The existing bounds only involve operator dimensions $d\leq 6$.
The possibility of helicity dependence is disregarded by many authors,
so it is also appropriate to set to zero the coefficients
$\gurfc{6}{}$ and $\Hurfc{5}{}$ in these cases.
The reported bounds involve a variety of particle species,
including electrons, muons, taus, protons, and quarks.
For the latter,
all the partonic quarks are assumed to have
the same dispersion relation.
Since the present focus is on fermions,
Lorentz violation in bosons such as pions or photons
is neglected for simplicity when making conversions.

Table \ref{tabletranslation} compiles 
some resulting constraints on isotropic Lorentz violation in the SME. 
The first two columns of this table
list the operator dimension and the sector of the SME involved.
The table includes constraints on minimal SME operators with $d=4$
as well as on ones with nonminimal dimensions.
The next three columns of the table contain the constraints on 
ultrarelativistic isotropic spherical coefficients
obtained from existing bounds.
The coefficients for different particle species 
are distinguished with a subscript denoting the species in question.
The final column provides the source from which the constraint is extracted.

The table reveals that the existing bounds span different coefficients.
However,
of the seven types of  
possible isotropic ultrarelativistic spherical coefficients 
with $d\leq 6$,
namely 
$\aurfc{3}{}$, $\curfc{4}{}$,  $\gurfc{4}{}$,  
$\aurfc{5}{}$,  $\Hurfc{5}{}$, $\curfc{6}{}$, and $\gurfc{6}{}$,
constraints exist on at most four of them for any one species.
We see that even within the very restrictive assumption
of isotropic ultrarelativistic Lorentz violation,
much of the coefficient space is unconstrained to date.
Also notably lacking are limits for neutral fermions,
including neutrons and other baryons.
We remark in passing that
numerous constraints exist on nonisotropic minimal fermion operators
\cite{tables},
including some extracted from studies of mesons
and some at impressive sensitivities.
Nonetheless,
the experimental coverage of SME coefficients in the fermion sector
is at present limited to a tiny fraction of the available possibilities.

\section{Summary}
\label{Summary}

In this work,
the general quadratic theory of a single Dirac fermion
in the presence of Lorentz violation
has been developed.
Our discussion began with the construction and basic properties
of the theory \rf{lag},
including two useful decompositions 
of the general spinor-matrix operator $\Qhat$ for Lorentz violation.
The first reveals the different spin content via the operators 
$\Shat$, $\Phat$, $\Vhat^\mu$, $\Ahat^\mu$, $\That^\mn$,
while the second displays CPT and other properties
via the notation
$\mhat$, $\mfivehat$, 
$\ahat^\mu$, $\bhat^\mu$, $\chat^{\mn}$, $\dhat^{\mn}$,
$\ehat^\mu$, $\fhat^\mu$, $\ghat^{\mu\rh\nu}$,
$\Hhat^\mn$
paralleling the conventions in the minimal SME. 
Table \ref{free_summary} compiles
some features of the corresponding coefficients for Lorentz violation.
In Sec.\ \ref{Field redefinitions},
we show that the physical observables in the pure quadratic theory \rf{lag}
are restricted to pieces of $\Vhat^\mu$ and $\That^\mn$,
generalizing known results for the minimal SME
and for the nonminimal neutrino sector.

We next constructed the exact dispersion relation 
for a fermion wave packet,
obtaining the closed and compact form \rf{exactdr}.
For some practical applications,
an approximate expression for the energy
valid at leading order in Lorentz violation
is useful,
and this is provided in Eq.\ \rf{drapprox2}.
The form of this equation
reveals that 
fermions experience anisotropy, dispersion, and birefringence
when in the presence of Lorentz violation.
The covariant projection operator 
yielding the spinor polarization is derived,
and the corresponding relativistic polarization vector
is given in Eq.\ \rf{polvec}.

With these key results in hand,
we next turned to the construction 
of the particle and antiparticle hamiltonians
associated with the theory \rf{lag}.
The 2$\times$2 hamiltonian for particles
is given as Eq.\ \rf{h} in Sec.\ \ref{Construction},
while that for antiparticles is in Eq.\ \rf{hbar}.
Using the relativistic polarization vector,
we can reduce the structure of these expressions
to the conceptually simple form \rf{pham},
which separates the particle hamiltonian
into four pieces according to spin and CPT properties.

Despite its conceptual simplicity,
the explicit form of the hamiltonian \rf{pham}
involves coefficients with numerous indices
and is unwieldy for many practical applications.
In Sec.\ \ref{Spherical decomposition},
we have taken advantage of the approximate rotation symmetry 
present in many experimental situations
to decompose the hamiltonian in spherical harmonics. 
The result \rf{helh} involves 
eight sets of spherical coefficients 
that characterize all types of Lorentz violation
for a single Dirac fermion.
Table \ref{spherical_coefs} summarizes
the basic properties of these coefficients.
Their comparatively simple properties under rotation, 
exemplified in Eq.\ \rf{coefrot},
make them well suited to explicit analyses.
The isotropic limit of the perturbative hamiltonian,
which can be useful in some treatments,
is obtained in Eq.\ \rf{phami},
and the corresponding isotropic Lagrange density
for operator dimensions $d=3,4,5,6$
is given in Eqs.\ \rf{three} through \rf{six}.

For many practical purposes,
limiting cases of the general formalism are useful.
Section \ref{Limiting cases}
extracts the nonrelativistic and ultrarelativistic cases 
and their isotropic limits.
The nonrelativistic hamiltonian is given in Eq.\ \rf{phamnr},
and the corresponding coefficients are summarized in
Table \ref{nr_coefs}.
The ultrarelativistic hamiltonian is presented in Eq.\ \rf{phamur},
and Table \ref{ur_coefs} lists properties of its coefficients.
This section also explicitly connects the spherical decomposition 
for operators of renormalizable dimension 
with standard expressions for the minimal SME.

The final technical discussions in this paper
concern immediate applications of our results.
In Sec.\ \ref{Dispersion and birefringence},
the issue of dispersion and birefringence
is revisited in the spherical language.
The dispersion relation, group velocity,
and the spin-precession rate \rf{sppr}
are derived in compact forms
in various limiting cases.
We then address in Sec.\ \ref{Connections to other formalisms}
the relationships between the present general framework
and some special field-theoretic models and kinematical approaches
in the literature.
The combination of the above results permits
translation of a wide variety of existing astrophysical bounds
on isotropic Lorentz violation
into constraints on isotropic spherical SME coefficients,
which are compiled in Table \ref{tabletranslation}.

Overall,
the results in this paper offer a comprehensive theoretical framework
for investigations of Lorentz and CPT violation 
involving quadratic fermion operators.
The physical effects identified here
provide a basis for future experimental searches.
Numerous types of Lorentz and CPT violation
are unconstrained to date,
and the prospects for exploration and
the potential for discovery remain bright.

\bigskip

This work was supported in part
by the Department of Energy
under grant DE-FG02-13ER42002
and by the Indiana University Center for Spacetime Symmetries.

\end{document}